\documentclass[prl,twocolumn,aps,showpacs]{revtex4-1}

\usepackage[TS1,OT1,T1]{fontenc}
\usepackage{hyperref}
\usepackage{graphicx}
\usepackage{sidecap}
\usepackage{verbatim}

\begin{document}
\title{Realizing the Harper Hamiltonian with Laser-Assisted Tunneling in Optical Lattices}
\author{Hirokazu Miyake, Georgios A. Siviloglou, Colin J. Kennedy, William Cody Burton, Wolfgang Ketterle}
\affiliation{MIT-Harvard Center for Ultracold Atoms, Research Laboratory of Electronics, Department of Physics, Massachusetts Institute of Technology, Cambridge, Massachusetts 02139, USA}
\date{\today}
\begin{abstract}
We experimentally implement the Harper Hamiltonian for neutral particles in optical lattices using laser-assisted tunneling and a potential energy gradient provided by gravity or magnetic field gradients. This Hamiltonian describes the motion of charged particles in strong magnetic fields. Laser-assisted tunneling processes are characterized by studying the expansion of the atoms in the lattice. The band structure of this Hamiltonian should display Hofstadter's butterfly. For fermions, this scheme should realize the quantum Hall effect and chiral edge states.
\end{abstract}
\maketitle

Systems of charged particles in magnetic fields have led to many discoveries in science -- including both the integer \cite{Klitzing1980prl} and the fractional \cite{FQHE1982,FQHE1983} quantum Hall effects -- and have become important paradigms of quantum many-body physics \cite{Lewenstein}. Generalizations have led to important developments in condensed matter physics, including topological insulators \cite{Kane2005prl,Konig2007}, fractional Chern insulators \cite{Regnault2011prx,Moller2009prl}, and Majorana fermions \cite{Fu2008prl,Mourik2012sci}.  At high magnetic fields, exotic new phenomena like the fractal energy spectrum of Hofstadter's butterfly \cite{Hofstadter:1976ys} are predicted to emerge. Its direct observation would require an inaccessibly high magnetic field of one flux quantum per unit cell -- corresponding to $\sim10,000$ Tesla in a traditional condensed matter system. Recently, some aspects of Hofstadter's butterfly were addressed using superlattices in high magnetic fields \cite{Hunt2013, Novoselov2013,Dean2013nature,vonKlitzing2001prl}.

Neutral atoms provide an excellent platform to simulate the physics of charged particles in magnetic fields free from disorder.  Rotating quantum gases realize the limit of weak magnetic fields, exploiting the equivalence between the Lorentz force and Coriolis force.  The observed vortex lattices \cite{Dalibard2000,AboShaeer2001} are analogous to magnetic flux lattices. A more general method to create synthetic magnetic fields for neutral atoms is based on the insight that vector potentials introduce spatially-varying phases $\phi$ into the wavefunction when the particle propagates, $\phi= \oint \mathbf{A}\cdot \mathbf{ds}/\hbar$, where the charge is included in the vector potential.  For neutral atoms, such phase structure can be realized through Berry phases, when two hyperfine states of the atom are coupled by Raman lasers with inhomogeneous intensity or detuning \cite{Lin2009,Dalibard2011rmp}.  This concept of coupling of two or several internal states to realize synthetic magnetic fields was also suggested in optical lattice geometries \cite{Jaksch:2003fk,Mueller2004pra,Gerbier2010njp}.  Here the crucial element is laser-assisted hopping between neighboring sites which imprints the phase of the laser into the atomic wavefunction.  Alternatively, instead of using Raman laser beams, lattice modulation techniques can generate complex tunneling matrix elements in optical lattices \cite{Struck2012, Struck:2013ly}. Experimentally, these techniques have been used so far only to realize staggered magnetic fields \cite{Aidelsburger2011, Struck:2013ly}. In the Munich experiment, the two internal states in the proposed schemes \cite{Jaksch:2003fk,Gerbier2010njp} were replaced by doubling the unit cell of the optical lattice using superlattices \cite{Aidelsburger2011}.

So far, all proposals for generating high magnetic fields are based on the coupling of different internal states. For alkali atoms, this involves different hyperfine states \cite{Jaksch:2003fk}. Spin flips between such states requires near-resonant light which heats up the sample by spontaneous emission. At least for staggered fluxes, the realizations with lattice shaking and superlattices demonstrates that internal structure of the atom is not essential.  Here we suggest and implement a scheme which realizes the Harper Hamiltonian \cite{Harper1955}, a lattice model for charged particles in magnetic fields, the spectrum of which is the famous Hofstadter's butterfly \cite{Hofstadter:1976ys}. Our scheme requires only far-off resonant lasers and a single internal state.  It is an extension of a scheme suggested by Kolovsky \cite{Kolovsky:2011bh}, which was shown to be limited to  inhomogeneous fields \cite{creffield2013epl}, but as we show here an additional momentum transfer in the laser-assisted hopping process provides a simple solution. While this work was in progress \cite{Miyake2013}, an identical scheme was proposed by the Munich group \cite{Aidelsburger:2013ys}.  In this paper, we describe the features and implementation of this scheme, and characterize the laser-assisted hopping process.

We start with the simple Hamiltonian for non-interacting particles in a 2D cubic lattice,
\begin{equation} \label{H_vanilla}
H=-\sum_{\langle m,n\rangle} \big(J_x \hat a^{\dagger}_{m+1,n}\hat a_{m,n} + J_y\hat a^{\dagger}_{m,n+1}\hat a_{m,n} + h.c. \big)
\end{equation}
where $J_{x(y)}$ describes tunneling in the  $x$-($y$-)direction and $\hat a^{\dagger}_{m,n}$ ($\hat a_{m,n}$) is the creation (annihilation) operator of a particle at lattice site $(m,n)$. Brackets indicate summation over only nearest-neighbor sites. Tunneling in the $x$-direction is then suppressed by a linear tilt of energy $\Delta$ per lattice site, where $\Delta/h$ is the Bloch oscillation frequency.  This tilt can be created with magnetic field gradients, gravity, or an AC Stark shift gradient. Resonant tunneling is restored with two far-detuned Raman beams of two-photon Rabi frequency $\Omega$, frequency detuning $\delta \omega = \omega_1 - \omega_2$, and momentum transfer $\delta \mathbf{k} = \mathbf{k_1} - \mathbf{k_2}$, as shown in Figure 1a.  Note that the two Raman beams couple different sites, but do not change the internal state of the atoms.  For resonant tunneling, $\delta \omega = \Delta/\hbar$, time-averaging over rapidly oscillating terms \cite{Jaksch:2003fk} yields an effective Hamiltonian which is time-independent. As a result, the tilt has disappeared because, in the dressed atom picture, site $(m,n)$ with $j$ and $k$ photons in the two Raman beams is degenerate with site $(m+1,n)$ and $j+1$ and $k-1$ photons in the two beams. This effective Hamiltonian describes the system well assuming that $\Delta$ is larger than the bandwidth, $\sim$$J$, and smaller than the bandgap, $E_{\text{Gap}}$.
\begin{figure}
\centering
\includegraphics[width=0.48\textwidth]{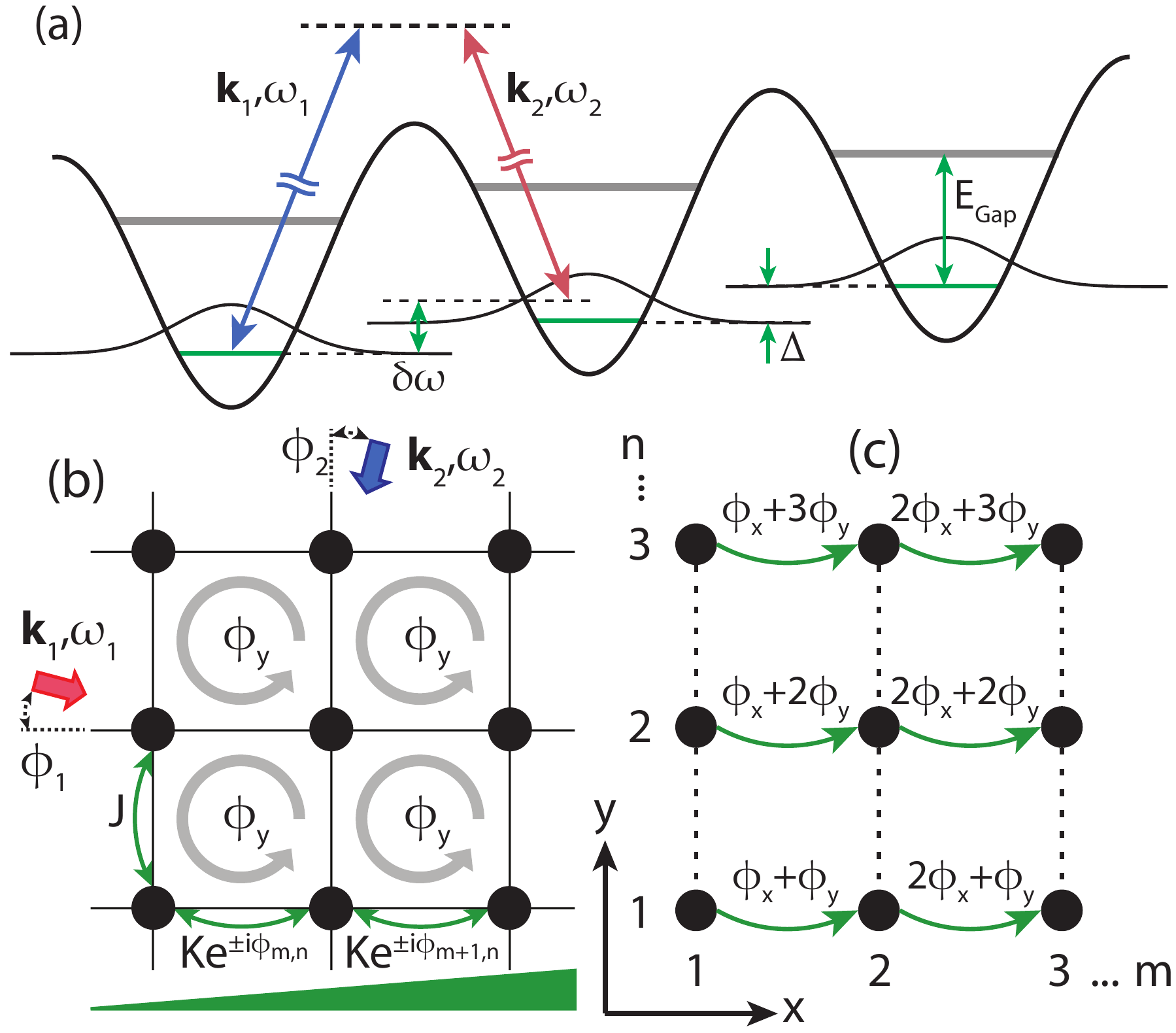}
\caption{(a) Raman-assisted tunneling in the lowest band of a tilted lattice with an energy offset $\Delta$ between neighboring sites. The two-photon Rabi frequency $\Omega$ determines the coupling between adjacent wells. (b) Experimental geometry to generate uniform magnetic fields using a pair of far-detuned laser beams and a uniform potential energy gradient. Tunneling along the x-direction with amplitude $K$ imprints a complex, spatially-varying phase $\phi_{m,n}$ -- with site indices ($m,n$) -- into the system due to the momentum transfer in the y-direction.  (c) A schematic depicting the position-dependent phases of the tunneling process. The equivalent number of flux quanta per unit cell is $\alpha=\phi_y/2\pi$.}
\end{figure}
The resulting Hamiltonian is equivalent to one that describes charged particles on a lattice in a magnetic field under the tight-binding approximation \cite{Harper1955,Hofstadter:1976ys} -- the single-band Harper Hamiltonian:
\begin{equation} \label{Harper}
H=-\sum_{\langle m,n\rangle} \big( Ke^{-i\phi_{m,n}}\hat a^{\dagger}_{m+1,n}\hat a_{m,n} + J\hat a^{\dagger}_{m,n+1}\hat a_{m,n} +h.c. \big)
\end{equation}
with spatially-varying phase, $\phi_{m,n}  = \delta \mathbf{k}\cdot\mathbf{R}_{m,n}= m\phi_x + n\phi_y$. Solutions in this model are periodic with respect to the number of flux quanta per unit cell, $\alpha$. If the frequency of the Raman beams are similar to those used for the optical lattice, one can tune $\alpha$ over the full range between zero and one by adjusting the angle between the Raman beams, and consequently $k_y$. A similar Hamiltonian can be realized for the tunneling of phonons between ion microtraps \cite{Bermudez2011prl}.

The spatially-dependent phase imprinted by the Raman lasers, given by $\phi_{m,n}$, can be intuitively understood in a pertubative regime where, $J=J_y$ and:
\begin{eqnarray} \nonumber
K&=&\frac{\Omega}{2}\int d^{2}\mathbf{r}\:w^{*}(\mathbf{r} - \mathbf{R}_{m,n})e^{-i\delta \mathbf{k} \cdot \mathbf{r}}w(\mathbf{r} - \mathbf{R}_{m,n}-a\hat \mathbf{x}) \\
&=& Ke^{-i\delta\mathbf{k}\cdot \mathbf{R}_{m,n}}
\end{eqnarray}
where $\mathbf{R}_{m,n}$ denotes the position of each lattice site.  Adding up the accumulated phases around a closed path, one sees that this method leads to an enclosed phase of $\phi_{y}=\delta k_y a$ per lattice unit cell of area $a^2$, thus realizing the Harper Hamiltonian with $\alpha=\phi_{y}/2 \pi$.

In a cubic lattice, the Wannier function $w(\mathbf{r})$ factorizes into $w(x)w(y)$ which are the localized Wannier-Stark and Wannier wavefunctions, respectively. The resulting expression for $K = \frac{\Omega}{2}\int dx\,w^{*}(x)e^{-i k_x x}w(x-a)\times \int dy\,w^{*}(y)e^{-i k_y y}w(y)$ shows that the momentum transfer in the $x$-direction is necessary to have a non-vanishing tunneling matrix element $K$. The $x$ momentum transfer does not contribute to the enclosed flux (or the value of the synthetic magnetic field B), but to the vector potential $\mathbf{A} =  \hbar(k_y y + k_x x)/a\: \hat{\mathbf{x}}$.  Therefore, our scheme does not realize the simple Landau gauge for the magnetic field. Note that it is this momentum transfer along the $x$-direction that distinguishes our scheme from Refs. \cite{Jaksch:2003fk,Gerbier2010njp,Kolovsky:2011bh}, and is responsible for connecting the two orthogonal Wannier states in the $x$-direction without changing the internal state.

For a more comprehensive description, we add the moving lattice -- $V_{RM} = \Omega \sin(\delta\mathbf{k}\cdot \mathbf{r} - \omega t)$ -- of the two Raman lasers along with a linear tilt to the Hamiltonian in Eq. \ref{H_vanilla}. In addition to the off-diagonal laser-assisted tunneling term, this moving lattice causes a diagonal term, which is a temporal modulation of the on-site energies.  A unitary transformation as in \cite{Tino2012,Miyake2013} leads to a frame rotating non-uniformly in time and position that eliminates the diagonal time-dependence.  For resonant drive, $\Delta = \hbar\delta\omega$, the onsite energies are all equal and vanish while the remaining off-diagonal coupling has a time-independent part leading to the Harper Hamiltonian as in Eq. \ref{Harper}. The resulting expressions for $K$ and $J$ due to the temporal modulation of the lattice and one-site wavefunction are (see supplemental information):
\begin{eqnarray} \label{BesselK} \nonumber
K&=& \Omega\Phi_{y0}e^{-i \phi_{m,n}} \bigg[ \Phi_{x1} \frac{J_{1}(\Gamma_x )}{\Gamma_x} + i \Phi'_{x1} \frac{dJ_{1}(\Gamma_x )}{d\Gamma_x} \bigg] \\ \label{BesselJ}
J&=&J_y J_{0}(\Gamma_y),\: \: \: \: \Gamma_i = \frac{2\Omega\Phi_{y0}\Phi_{x0}}{\Delta}\sin\bigg(\frac{k_i a}{2}\bigg)
\end{eqnarray}
where $\Phi_{i0} = \langle 0|\cos(k_i x_i)| 0\rangle$ is the on-site matrix element, and $\Phi_{x1} = \langle 0|\sin(k_x (x-a/2))| 1\rangle$ and $\Phi'_{x1} = \langle 0|\cos(k_x (x-a/2))| 1\rangle$ are the off-diagonal matrix elements. This result is more general than the case of phase modulation \cite{Tino2012} and the tight-binding limit in \cite{Aidelsburger:2013ys,Aidelsburger2013arxiv}, where $K$ is proportional to $J_1 (x)$. 

We implement the Harper Hamiltonian with each Raman laser aligned along one of the two lattice directions, $x$ and $y$, corresponding to momentum transfer in both directions of $\hbar k_L$ -- the single photon recoil of the lattice laser. The magnetic flux per unit cell resulting from $k_y = k_L$ is $\alpha = 1/2$. In the tight-binding limit for this momentum transfer, $\Phi_{i0} \approx 1$ and $\Phi_{x1} \approx -2J_x/\Delta \gg \Phi'_{x1}$, so the resonant tunneling amplitudes resulting from $k_x = k_L$ simplify to: 
\begin{equation} \label{Bessels}
K = J_x J_{1} \bigg( \frac{2\Omega}{\Delta}\bigg) \text{, and: } J = J_y J_{0} \bigg( \frac{2\Omega}{\Delta}\bigg)
\end{equation}

Experimentally, the system is prepared by starting with a Bose-Einstein condensate of $\sim 5 \times 10^5$ $^{87}$Rb atoms in the $|2,-2\rangle$ state in a crossed dipole trap. The Raman lasers are ramped up to their final intensities in 30 ms at a large detuning of $200$ kHz, far away from any excitations of the system, and are switched to their final detuning after the tilt is applied to the system (see below). To avoid interference between the lattice and Raman lasers, they are perpendicularly polarized and frequency offset by $>$50 MHz using acousto-optic modulators. Next, we adiabatically load the condensate in 100 ms into a two-dimensional cubic optical lattice of spacing $\lambda_{latt}/2=532$ nm. For longer hold times, a weak 2 $E_r$ lattice beam along the third direction is simultaneously ramped up to provide additional confinement.  Here, $E_r=\hbar^2 k_{L}^2 \big/ {2m}\approx$ $h\times$2 kHz is the single photon recoil. Lattice depths are calibrated using Kapitza-Dirac scattering, and the two photon Rabi frequency of the Raman lasers is determined using free-space Rabi oscillations.

After loading the condensate into the lattice, a uniform potential energy gradient is applied by turning off the confining crossed dipole traps in $20$ ms. This exposes the cloud to a linear gravitational potential (which was compensated until then by the trapping beams).  Alternatively, we have successfully used a magnetic field gradient to access a broader range of tilts. The data presented here were obtained with the gravitational force which provides an offset of $ mga/h\approx 1.1$ kHz between adjacent lattice sites. This has the advantage over the magnetic gradient of a much faster switching time.  The cloud widths, $\sigma_x$ and $\sigma_y$ are obtained by standard absorption imaging along the direction perpendicular to the 2D lattice.

\begin{figure}
\centering
\includegraphics[width=0.48\textwidth]{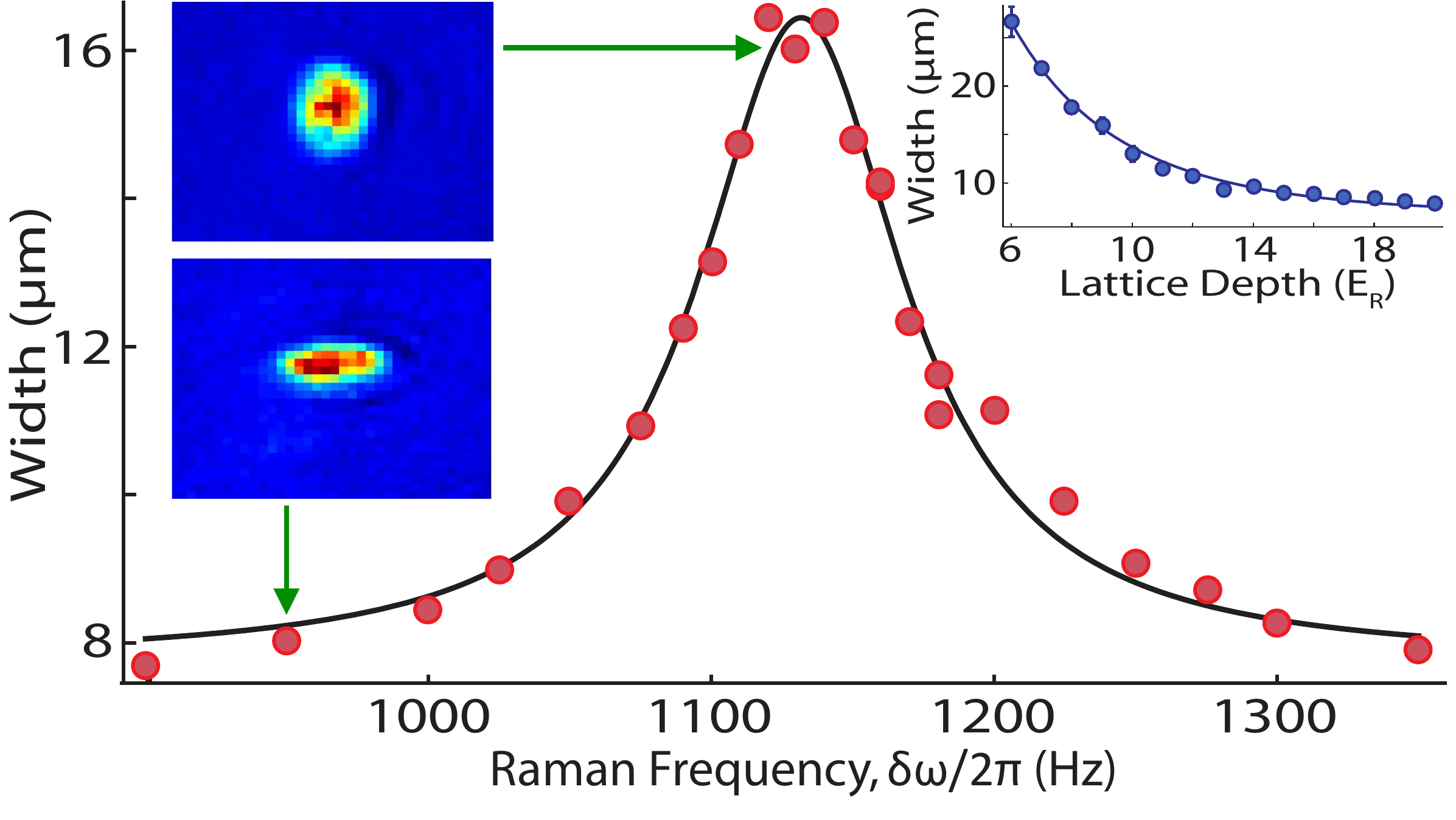}
\caption{\textit{In situ} cloud width as a function of Raman detuning, $\delta \omega$, after an expansion of 500 ms, with a Raman lattice depth of $\Omega = \Delta/4$. The solid line is a Lorentzian fit to the experimental data (dots) centered at 1133 Hz -- consistent with the gravitational offset between sites. Pictures (of size $135\times$116 $\mu$m) show typical column densities on/off resonance. (Inset) Dependence of the laser-assisted tunneling on optical lattice depth. For deeper lattices, the expansion occurs more slowly.}
\end{figure}

The essential feature of our implementation of the Harper Hamiltonian is that tunneling in the $x$-direction is suppressed by a potential tilt, and reestablished by laser-assisted tunneling.  This is demonstrated in Figure 2 which shows the resonance for the laser-assisted process.  For this, tunneling is characterized by looking at the expansion of the cloud within the lattice. Expansion occurs since the confinement by the optical dipole trap has been switched off, and due to some heating during the $500$ ms hold time.  Note that for fully coherent time evolution, charged particles in a magnetic field will undergo cyclotron motion which would suppress the expansion.  The resonance width of $60$ Hz may have contributions from laser frequency jitter, inhomogeneous lattice potential and atomic interactions.  The Lorentzian fit suggests a homogenous broadening mechanism. Fig. 2 demonstrates how the laser-assisted tunneling rate can be controlled by the lattice depth.

\begin{figure}
\centering
\includegraphics[width=0.48\textwidth]{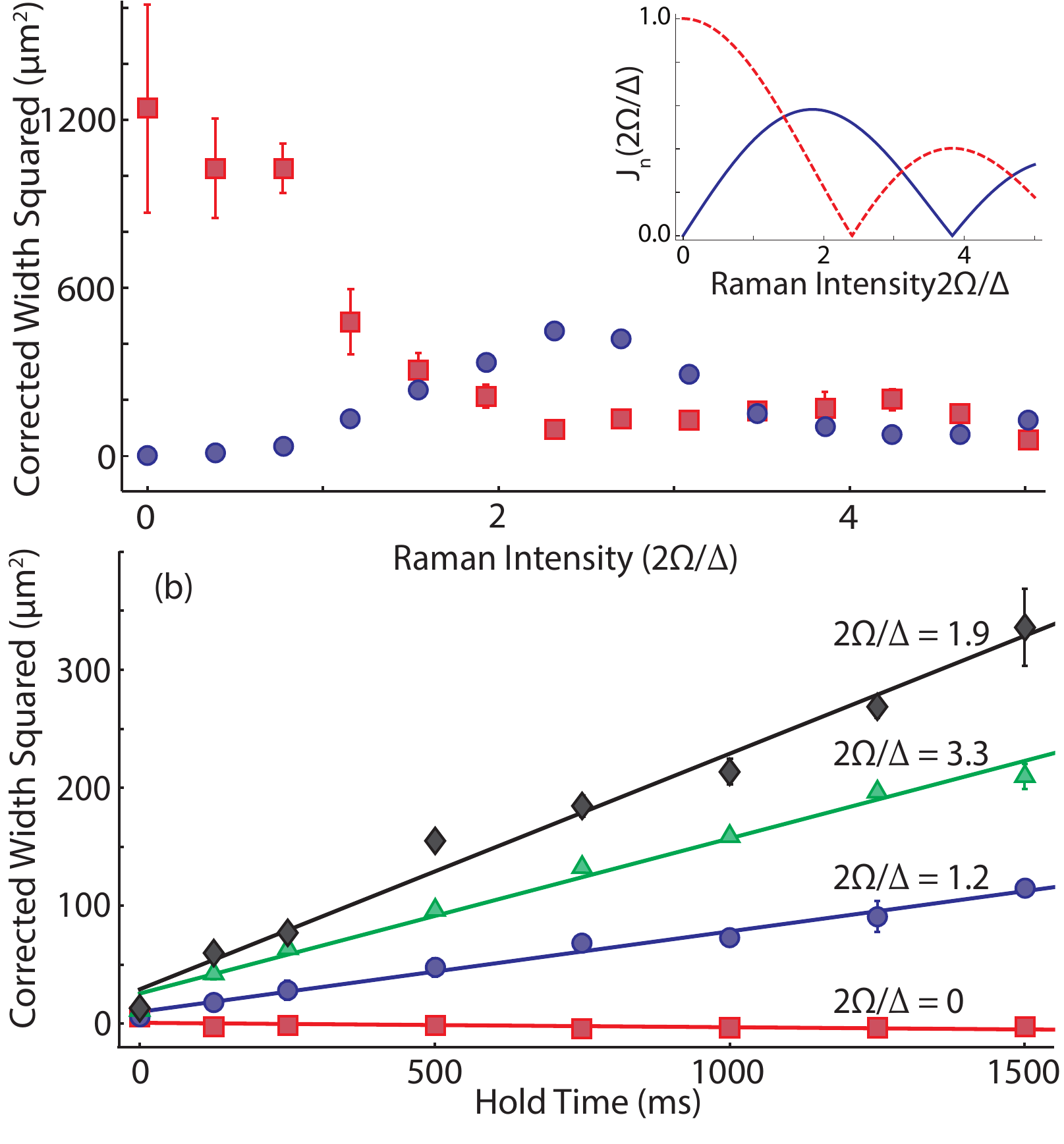}
\caption{(a)Expansion as a function of resonant Raman laser intensity shows the laser-assisted tunneling along the tilt direction (blue), and the tunneling rate $J$ along the transverse direction (red). Data taken at lattice depths of 9 $E_r$ and hold time of 1500 ms. (Inset) Theoretical prediction for the tunneling rates $K$ and $J$ in terms of Bessel functions (Eq. \ref{Bessels}).  (b) Time evolution of the squared width for different Raman laser intensities.   From the slope of the lines, we obtain the laser-assisted tunneling rates and their statistical errors: $0.2\pm0.08$ (red squares), $4\pm0.5$ (blue circles), $12\pm1$ (black diamonds), and $8\pm0.5$ Hz (blue triangles).}
\end{figure}

The dependence of $K$ and $J$ on the intensity of the Raman lasers (described by Bessel functions) allows tuning of the ratio of the two.  For low intensities, $K$ increases linearly with the intensity, and $J$ decreases quadratically. The latter reflects the depletion of the unperturbed Wannier function by the modulation due to the moving Raman lattice.  Fig. 3a shows experimental results in qualitative agreement with these predictions.

For a quantitative interpretation of the expansion of the cloud, we assume an incoherent diffusion process, where the square of the width $\sigma$ of the expanded cloud is proportional to the tunneling rate times expansion time.  For finite time, we correct for the initial size $\sigma_{0}$ by assuming that the expansion and initial size add in quadrature, and plot the corrected squared width $\sigma_{corr}^{2}=\sigma^{2}- \sigma_{0}^{2}$ versus time.  The slope is proportional to the laser-assisted tunneling rate. Absolute tunneling rates are obtained by comparing this result to the expansion of the cloud in the y-direction with the Raman beams far off resonance, when normal tunneling occurs.  The ratio of the slopes is then $K/J_y$, with $J_y$ calculated from the calibrated lattice depth to be $\sim$ $h \times 48\,\text{Hz}$.  Figure 3b shows the time evolution of the square of the corrected size for various Raman intensities. The linear fits supports the assumption of incoherent diffusion and allows a determination of tunneling rates as summarized in the figure caption.

Laser-assisted tunneling is a powerful tool to manipulate the motion of atoms in optical lattices and to create novel Hamiltonians.  We now describe different tunneling processes observed by a wide scan of the Raman detuning, shown in Figure 4. A strong peak near 568 Hz fulfills the resonance condition $2\delta\omega=\Delta/\hbar$ for a four-photon nearest-neighbor tunneling process.  This resonance is similar to the one observed in Ref. \cite{Sias2008} by shaking the lattice.  Note that the four-photon resonance is narrower (20 Hz versus 60 Hz) than the two-photon resonance, indicative of a higher-order process. Broad features at even lower frequency are most likely due to higher order tunneling resonances and low-lying excitations within the first band.

Next-nearest-neighbor tunneling occurs at  $\delta\omega=2\Delta/\hbar$, twice the frequency of the fundamental resonance. For a shaken lattice (no Raman beams) this was studied in Ref. \cite{Ivanov2008}.  Analyzing the expansion of the cloud gives a tunneling rate of $0.4\pm0.1$ Hz, comparable to the next-nearest-neighbor tunneling rate in an untilted lattice, $\sim 0.8$ Hz in our system.  However, in an untilted lattice, next-nearest neighbor tunneling is typically hundred times slower that nearest neighbor tunneling, whereas in laser-assisted tunneling, the two processes can be independently controlled by the laser power at the two resonant frequencies.  Tunneling rates below 1 Hz are too slow for pursuing many-body physics, but the same scheme can be implemented for lighter atoms such as lithium in a shorter wavelength lattice, where the relevant scale factor, the recoil energy, is increased by a factor of fifty.
 
\begin{figure}
\centering
\includegraphics[width=0.48\textwidth]{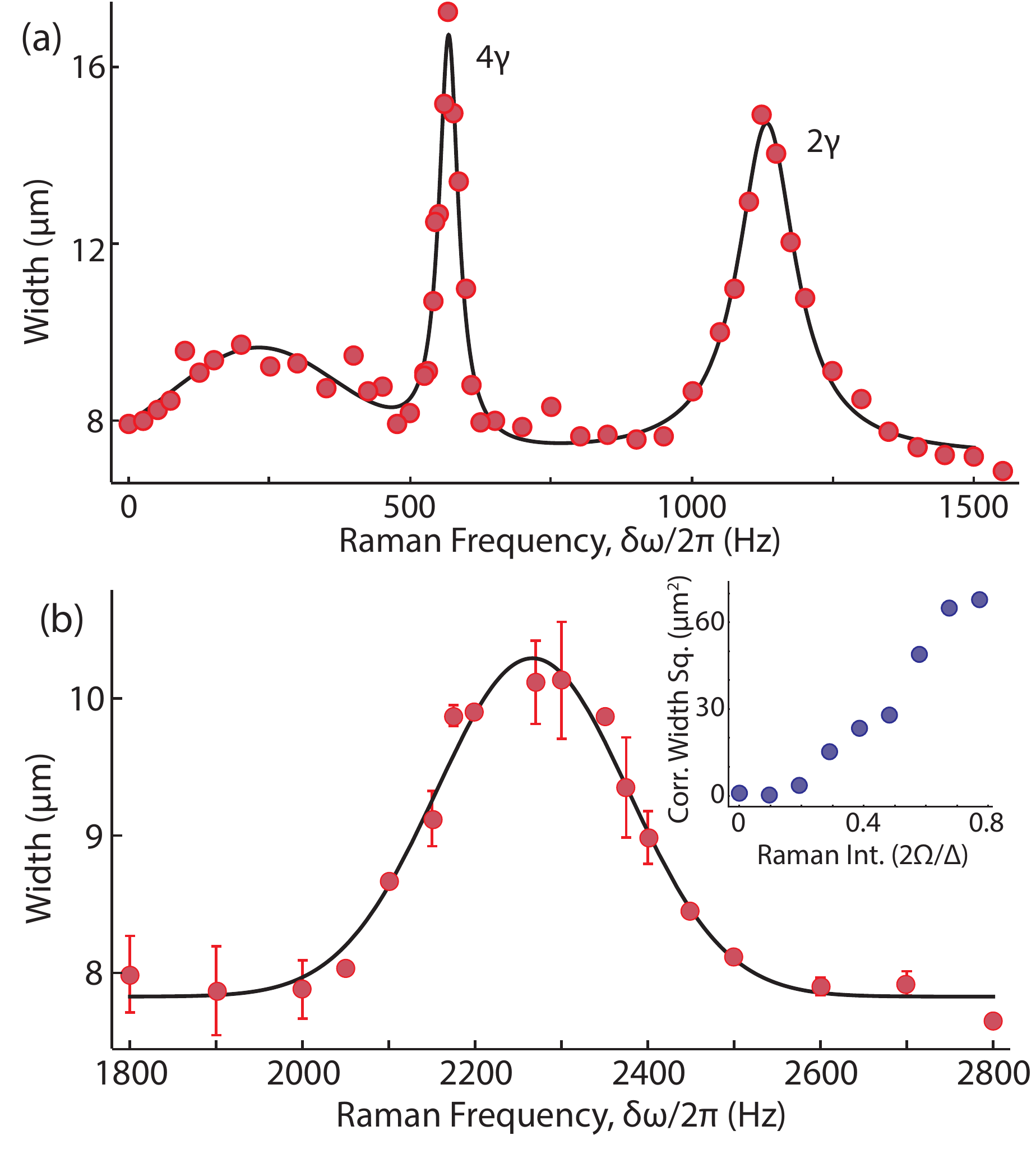}
\caption{Spectrum of excitations and tunneling resonances. (a) A strong, four photon, nearest neighbor tunneling resonance appears at $\Delta/2h$ along with the $K$ resonance at $\Delta/h$. This data was taken at a lattice depth of 9 $E_r$, for a two-photon Raman Rabi frequency of 1092 Hz and  500 ms expansion time. (b) Observation of next-nearest-neighbor laser-assisted tunneling at $2\Delta$. (Insert) Expansion with $\Omega$, no saturation is reached. Experimental conditions as in (a), except for an expansion time of 1500 ms.}
\end{figure}

After realizing and characterizing all parts of the Harper Hamiltonian, the next goal is to map out its band structure as a function of quasimomentum and magnetic field $\alpha$ -- the Hofstadter butterfly.  The ground state for a given $\alpha$ should be accessible by adiabatically transferring a condensate into this Hamiltonian.  The ground state of the Harper Hamiltonian for $\alpha=1/2$ has a clear signature in that its magnetic unit cell is twice as large as the lattice unit cell and its wavefunction has a unit cell that is four times as large, so time of flight imaging will reveal the resulting reduction of the Brillouin zone in momentum space by a factor of four \cite{Moller2010pra,Aidelsburger2011,Miyake2013,Polak2013,Powell2011pra}.  So far, we have not been able to preserve the low entropy of the initial condensate.  

Preliminary experiments have shown that there is less heating by the Raman beams at larger frequency detunings requiring larger magnetic field gradients. Another potential source of heating is atomic interactions. Instabilities of certain quasi-momentum states in optical lattices have been studied in Refs. \cite{Fallani2004,Campbell2006}.  Interaction-induced heating effects can in principle be avoided by using Feshbach resonances to tune the scattering length to zero or by using a single spin component of a fermionic gas.  Once the ground state of the Harper Hamiltonian is established, different quasi-momentum states can be populated through Bloch oscillations which occur at frequency  $\delta = \delta \omega - \Delta/\hbar$, when the Raman lasers are sightly detuned from the resonance studied here.

The Harper Hamiltonian established in this work will be the starting point for many exciting explorations including the quantum Hall effect, Dirac points, and novel topological phenomena \cite{Moller2009prl,DasSarma2011}.  Interactions between atoms may also lead to bosonic Laughlin states \cite{Lukin2005prl}.  The lowest band is topologically non-trivial with a Chern number of one \cite{Lukin2007pra} and should show chiral edge states.  Most importantly, our scheme is simpler and potentially more robust than other suggestions, since it does not require near resonant light for connecting hyperfine states. It can be implemented for any atom -- including the workhorse fermionic atoms lithium and potassium -- which have small fine structure splittings, making it impossible to couple different spin states with negligible heating by spontaneous emission.

We acknowledge Yuri Lensky for experimental assistance. We thank Wujie Huang for helpful discussions and critical reading of the manuscript. This work was supported by the NSF through the Center of Ultracold Atoms, by NSF award PHY-0969731, under ARO Grant No. W911NF-13-1-0031 with funds from the DARPA OLE program, and by ONR.  During the writing of the manuscript, we became aware of similar work carried out in the group of I. Bloch in Munich \cite{Aidelsburger2013arxiv}.

\onecolumngrid
\section{Supplementary Material}

We start with the Wannier-Stark Hamiltonian and add a moving lattice to drive laser-assisted tunneling processes, which is described by a Raman process between states on different lattice sites. The derivation generalizes the treatment in \cite{Tino2012a,Miyake2013a}, and the notation utilizes that of the main paper.
\begin{equation}
H = \frac{\mathbf{p}^2}{2m} + V_{\text{latt}}(\mathbf{r}) - \frac{\Delta}{a} \mathbf{x} + \Omega \sin \Big( \delta\mathbf{k}\cdot\mathbf{r} - \frac{k_x a}{2} - \omega t \Big)
\end{equation}
The phase term, $-k_x a/2$, is included for later computational convenience, and does not change the physics of the problem. Working in two dimensions and ignoring interactions, the Wannier-Stark Hamiltonian is projected onto the lowest band of a cubic lattice using a basis of localized Wannier-Stark functions in the $x$-direction and Wannier functions in the $y$-direction:
\begin{eqnarray} \nonumber
H &=& \sum_{m,n} \Big( -m\Delta |m,n\rangle\langle m,n | - J_{y}|m,n+1\rangle \langle m,n| + h.c.\\ \nonumber
\ &+& \sum _{m',n'}\Omega |m',n'\rangle \langle m',n'|\sin \Big( \delta\mathbf{k}\cdot\mathbf{r} -\frac{k_x a}{2} - \omega t \Big)|m,n\rangle \langle m,n| \Big)
\end{eqnarray}

There are two matrix elements of interest here: the diagonal term as well as overlap of adjacent sites in the $x$- and $y$-directions. In general these have the form:
\begin{equation} \label{elements}
\langle m,n|\sin\Big(\delta\mathbf{k}\cdot\mathbf{r} -\frac{k_x a}{2} - \omega t\Big)|m+l,n+p\rangle
\end{equation}
The phase shift of the Raman drive, $k_x a/2$, is associated with a spatial shift for the tunneling matrix elements and a temporal shift for the on-site matrix elements making their self-consistent evaluation based on symmetry arguments easier. Using $\mathbf{R}_{m,n} = ma\hat{x} + na\hat{y}$ for the position of the lattice sites, the relevant matrix elements can be re-written as:
\begin{equation}
\langle 0,0|\sin\big(\delta\mathbf{k}\cdot(\mathbf{r}+\mathbf{R}_{m,n}) - k_x a/2 - \omega t\big)|l,p\rangle
\end{equation}
To condense notation, we define: $\theta_{m,n} = \omega t - \delta\mathbf{k}\cdot\mathbf{R}_{m,n} = \omega t - \phi_{m,n}$, with $\phi_{m,n} = mk_x a + nk_y a$. Expanding the $\sin(a+b-c)$ form of the Raman operator into four terms one obtains the relevant matrix elements:
\begin{eqnarray}
\langle 0|\sin(k_y y)| p=0\rangle &=& 0 \\
\langle 0|\cos(k_y y)| p=0\rangle &=& \Phi_{y0}(k_y) \\
\langle 0|\sin(k_x x)| l=0\rangle &=& 0 \\
\langle 0|\cos(k_x x)| l=0\rangle &=& \Phi_{x0}(k_x) \\
\langle 0|\sin(k_x (x -a/2))|l=1\rangle &=& \Phi_{x1}(k_x) \\
\langle 0|\cos(k_x (x -a/2))|l=1\rangle &=& \Phi'_{x1}(k_x)
\end{eqnarray}
The expressions above are evaluated using maximally localized Wannier functions in the $y$-direction and Wannier-Stark wavefunctions in the $x$-direction. Due to the symmetric nature of the localized Wannier function \cite{Kohn}, all matrix elements of an antisymmetric function are zero. The Wannier-Stark wavefunctions do not have definite parity as discussed in \cite{Gluck} so the overlap elements must be individually evaluated. In the tight-binding limit, the tunneling term is dominated by $\Phi_{x1} \approx -2J_x \sin(k_x a/2)/\Delta$. However, at lower lattice depths $\Phi'_{x1}$ can become significant so we keep both terms.

The coupling between adjacent sites in the tilted, $x$-direction becomes:
\begin{equation} 
\Omega\Phi_{y0} \big( \Phi_{x1}\cos\theta_{m,n} - \Phi'_{x1}\sin\theta_{m,n} \big)
\end{equation}
In addition, the Raman-coupling induces an on-site modulation given by:
\begin{equation}
-\Omega\Phi_{x0}\Phi_{y0}\sin(\theta_{m,n} + k_x a/2)
\end{equation}

Given the above form for the on- and off-diagonal terms, we arrive at an effective Hamiltonian:
\begin{eqnarray} \label{Hprime} \nonumber
H &=& \sum_{m,n} \Big[ \big(-m\Delta -\Omega\Phi_{x0}\Phi_{y0}\sin(\theta_{m,n} + k_x a/2)\big)|m,n\rangle \langle m,n| \ldots \\
&& \ldots + \Omega\Phi_{y0} \big( \Phi_{x1}\cos\theta_{m,n} - \Phi'_{x1}\sin\theta_{m,n} \big) |m+1,n\rangle \langle m,n| - J_y|m,n+1\rangle \langle m,n| + h.c. \big) \Big]
\end{eqnarray}
The time dependence of the diagonal terms can be eliminated via a unitary transformation into a rotating frame given by:
\begin{equation} \label{transform} \nonumber
U = \exp\Big[i\sum_{m,n} \bigg(m\omega t - \frac{\Omega\Phi_{x0}\Phi_{y0}}{\hbar \omega}\cos\bigg(\theta_{m,n} + \frac{k_x a}{2} \bigg)\bigg)|m,n\rangle \langle m,n|\Big] = \sum_{m,n} e^{i\Lambda_{m,n}} |m,n\rangle \langle m,n|
\end{equation}
where $\Lambda_{m,n} =m\omega t - \frac{\Omega\Phi_{x0}\Phi_{y0}}{\hbar \omega}\cos(\theta_{m,n} + k_x a/2)$. In this frame the Hamiltonian becomes $H' = U^{\dagger}HU - i\hbar U^{\dagger} (dU/dt)$. For the case of resonant drive where $\hbar \omega = \Delta$, the diagonal terms are zero, leaving only off-diagonal elements for tunneling in the $x$- and $y$-directions. Considering tunneling in the $x$-direction first and using the full expression for $\theta_{m,n}$, the exponential factor is:
\begin{eqnarray} \nonumber
e^{-i(\Lambda_{m+1,n} - \Lambda_{m,n})} &=& \exp \Big[-i\Big(\omega t - \frac{\Omega\Phi_{x0}\Phi_{y0}}{\hbar \omega} \Big(\cos \Big(\omega t + \frac{k_x a}{2} - \phi_{m+1,n}\Big) - \cos \Big( \omega t + \frac{k_x a}{2}- \phi_{m,n}\Big )\Big)\Big)\Big] \\
&=& \exp \Big[-i\Big(\omega t - \frac{2\Omega\Phi_{x0}\Phi_{y0}}{\hbar \omega} \Big(\sin \Big(\frac{k_x a}{2}\Big) \sin ( \omega t - \phi_{m,n}) \Big)\Big)\Big]
\end{eqnarray}
For tunneling in the $y$-direction, the transformation into the rotating frame adds the exponential factor:
\begin{eqnarray} \nonumber
e^{-i(\Lambda_{m,n+1} - \Lambda_{m,n})} &=& \exp \Big[i\frac{\Omega\Phi_{x0}\Phi_{y0}}{\hbar \omega} \Big(\cos (\omega t +\frac{k_x a}{2} - \phi_{m,n+1}) - \cos(\omega t + \frac{k_x a}{2}- \phi_{m,n})\Big)\Big] \\
&=& \exp \Big[i\frac{2\Omega\Phi_{x0}\Phi_{y0}}{\hbar \omega} \Big(\sin \Big( \frac{k_y a}{2} \Big) \sin \Big( \omega t + \frac{(k_x - k_y) a}{2} - \phi_{m,n}\Big)\Big)\Big]
\end{eqnarray}
If we define the quantity $\Gamma_{x(y)} = \frac{2\Omega\Phi_{x0}\Phi_{y0}}{\hbar \omega}\sin(\frac{k_{x(y)} a}{2})$ and use the Jacobi-Anger identity, $e^{ix\sin(\theta)} = \sum_{r} J_{r}(x)e^{ir\theta}$, both above expressions can be simplified to:
\begin{eqnarray}
e^{-i(\Lambda_{m+1,n} - \Lambda_{m,n})} &=& e^{-i\omega t} \sum_{r=-\infty}^{\infty} J_{r}(\Gamma_x ) e^{ir(\omega t- \phi_{m,n} )} \\
e^{-i(\Lambda_{m,n+1} - \Lambda_{m,n})} &=& \sum_{r=-\infty}^{\infty}  J_{r}(\Gamma_y ) e^{ir(\omega t +(k_x - k_y) a/2 - \phi_{m,n} )}
\end{eqnarray}
for $x$ and $y$ tunneling, respectively. Here, $J_{r}(\Gamma)$ are the Bessel functions of the first kind. Now the time-dependent tunneling amplitudes $K(t) = \langle m+1,n| H' |m,n\rangle$ and $J(t) = \langle m,n+1| H' |m,n\rangle$ in the rotating frame are given by:
\begin{eqnarray}
K(t) &=& \Omega\Phi_{y0} \big( \Phi_{x1}\cos(\omega t - \phi_{m,n}) - \Phi'_{x1}\sin(\omega t - \phi_{m,n}) \big) e^{-i\omega t}  \sum_{r} J_{r}(\Gamma_x ) e^{ir(\omega t - \phi_{m,n} )} \\
J(t) &=& -J_y \sum_{r} J_{r}(\Gamma_x ) e^{ir(\omega t + (k_x-k_y) a/2 - \phi_{m,n} )}
\end{eqnarray}
So far we have transformed Eq. \ref{Hprime} exactly. Time averaging over a period of $\tau\sim1/\Delta$ gives tunneling rates for the effective Hamiltonian, $H_{\text{eff}} = \langle H' \rangle_{\tau}$:
\begin{eqnarray} \label{K}
K_{\text{eff}} &=& \frac{\Omega\Phi_{y0}}{2}e^{-i \phi_{m,n}} \bigg[ \Phi_{x1} \bigg(J_{0}(\Gamma_x ) + J_{2}(\Gamma_x )\bigg) + i \Phi'_{x1} \bigg(J_{0}(\Gamma_x ) - J_{2}(\Gamma_x )\bigg) \bigg] = K e^{-i \phi_{m,n}} \\
J_{\text{eff}} &=& -J_y J_0 (\Gamma_y) = J
\end{eqnarray}
Using the Bessel function identities $\frac{1}{2}(J_0 (x) + J_2 (x)) = J_1(x) /x$ and $\frac{1}{2}(J_0 (x) - J_2 (x)) = \frac{dJ_1(x)}{dx}$, the above expression for $K$ simplifies to:
\begin{equation}
K= \Omega\Phi_{y0}e^{-i \phi_{m,n}} \bigg[ \Phi_{x1} \frac{J_{1}(\Gamma_x )}{\Gamma_x} + i \Phi'_{x1} \frac{dJ_{1}(\Gamma_x )}{d\Gamma_x} \bigg]
\end{equation}

These constitute the coefficients of an effective Hamiltonian in the rotating frame exactly analogous to the Harper Hamiltonian:
\begin{equation}
H_{\text{eff}} = \sum_{\langle m,n \rangle} (K e^{-i \phi_{m,n}} |m+1,n \rangle \langle m,n| + J |m,n+1 \rangle \langle m,n| + h.c. )
\end{equation}

Within the tight-binding model, where $\Phi_{x1} \approx -2J_x\sin (k_x a / 2)/\Delta \gg \Phi'_{x1}$ and $\Phi_{y0} = \Phi_{x0} \approx 1$, the expression for the tunneling amplitude in the $x$-direction becomes:
\begin{eqnarray}
K \approx -J_x J_{1}\bigg(\frac{2\Omega}{\Delta}\sin \bigg(\frac{k_x a}{2}\bigg)\bigg) =-J_x J_1 \bigg(\frac{2\Omega}{\Delta}\bigg)
\end{eqnarray}
where the last equality is for the specific case where $k_x a = k_y a = \pi$, as in our experiment.

Furthermore, in the limit of low Raman lattice depths, $\Gamma_x \ll 1$, where $J_0 (\Gamma_x) \approx 1 \gg J_2(\Gamma_x)$, Eq. \ref{K} reduces to the perturbative expression for laser-assisted tunneling given in the main text:
\begin{eqnarray}
K_{\text{eff}} &=& \frac{\Omega \Phi_{y0}}{2}e^{-i \phi_{m,n}}\Big( \Phi_{x1} + i \Phi'_{x1} \Big) = i \frac{\Omega}{2} e^{-i \phi_{m,n}} \langle m=0 | e^{-i k_x x} |m=1 \rangle\langle n=0 | \cos(k_y y) |n=0 \rangle \\
&=& i \frac{\Omega}{2} e^{-i \phi_{m,n}} \langle 0,0 | e^{-i \delta \mathbf{k}\cdot \mathbf{r}} |1,0 \rangle
\end{eqnarray}


\begin{thebibliography}{43}%
\makeatletter
\providecommand \@ifxundefined [1]{%
 \@ifx{#1\undefined}
}%
\providecommand \@ifnum [1]{%
 \ifnum #1\expandafter \@firstoftwo
 \else \expandafter \@secondoftwo
 \fi
}%
\providecommand \@ifx [1]{%
 \ifx #1\expandafter \@firstoftwo
 \else \expandafter \@secondoftwo
 \fi
}%
\providecommand \natexlab [1]{#1}%
\providecommand \enquote  [1]{``#1''}%
\providecommand \bibnamefont  [1]{#1}%
\providecommand \bibfnamefont [1]{#1}%
\providecommand \citenamefont [1]{#1}%
\providecommand \href@noop [0]{\@secondoftwo}%
\providecommand \href [0]{\begingroup \@sanitize@url \@href}%
\providecommand \@href[1]{\@@startlink{#1}\@@href}%
\providecommand \@@href[1]{\endgroup#1\@@endlink}%
\providecommand \@sanitize@url [0]{\catcode `\\12\catcode `\$12\catcode
  `\&12\catcode `\#12\catcode `\^12\catcode `\_12\catcode `\%12\relax}%
\providecommand \@@startlink[1]{}%
\providecommand \@@endlink[0]{}%
\providecommand \url  [0]{\begingroup\@sanitize@url \@url }%
\providecommand \@url [1]{\endgroup\@href {#1}{\urlprefix }}%
\providecommand \urlprefix  [0]{URL }%
\providecommand \Eprint [0]{\href }%
\providecommand \doibase [0]{http://dx.doi.org/}%
\providecommand \selectlanguage [0]{\@gobble}%
\providecommand \bibinfo  [0]{\@secondoftwo}%
\providecommand \bibfield  [0]{\@secondoftwo}%
\providecommand \translation [1]{[#1]}%
\providecommand \BibitemOpen [0]{}%
\providecommand \bibitemStop [0]{}%
\providecommand \bibitemNoStop [0]{.\EOS\space}%
\providecommand \EOS [0]{\spacefactor3000\relax}%
\providecommand \BibitemShut  [1]{\csname bibitem#1\endcsname}%
\let\auto@bib@innerbib\@empty
\bibitem [{\citenamefont {Klitzing}\ \emph {et~al.}(1980)\citenamefont
  {Klitzing}, \citenamefont {Dorda},\ and\ \citenamefont
  {Pepper}}]{Klitzing1980prl}%
  \BibitemOpen
  \bibfield  {author} {\bibinfo {author} {\bibfnamefont {K.~von}\ \bibnamefont
  {Klitzing}}, \bibinfo {author} {\bibfnamefont {G.}~\bibnamefont {Dorda}}, \
  and\ \bibinfo {author} {\bibfnamefont {M.}~\bibnamefont {Pepper}},\ }\href
  {\doibase 10.1103/PhysRevLett.45.494} {\bibfield  {journal} {\bibinfo
  {journal} {Phys. Rev. Lett.}\ }\textbf {\bibinfo {volume} {45}},\ \bibinfo
  {pages} {494} (\bibinfo {year} {1980})}\BibitemShut {NoStop}%
\bibitem [{\citenamefont {Tsui}\ \emph {et~al.}(1982)\citenamefont {Tsui},
  \citenamefont {Stormer},\ and\ \citenamefont {Gossard}}]{FQHE1982}%
  \BibitemOpen
  \bibfield  {author} {\bibinfo {author} {\bibfnamefont {D.~C.}\ \bibnamefont
  {Tsui}}, \bibinfo {author} {\bibfnamefont {H.~L.}\ \bibnamefont {Stormer}}, \
  and\ \bibinfo {author} {\bibfnamefont {A.~C.}\ \bibnamefont {Gossard}},\
  }\href {\doibase 10.1103/PhysRevLett.48.1559} {\bibfield  {journal} {\bibinfo
   {journal} {Phys. Rev. Lett.}\ }\textbf {\bibinfo {volume} {48}},\ \bibinfo
  {pages} {1559} (\bibinfo {year} {1982})}\BibitemShut {NoStop}%
\bibitem [{\citenamefont {Laughlin}(1983)}]{FQHE1983}%
  \BibitemOpen
  \bibfield  {author} {\bibinfo {author} {\bibfnamefont {R.~B.}\ \bibnamefont
  {Laughlin}},\ }\href {\doibase 10.1103/PhysRevLett.50.1395} {\bibfield
  {journal} {\bibinfo  {journal} {Phys. Rev. Lett.}\ }\textbf {\bibinfo
  {volume} {50}},\ \bibinfo {pages} {1395} (\bibinfo {year}
  {1983})}\BibitemShut {NoStop}%
\bibitem [{\citenamefont {Lewenstein}\ \emph {et~al.}(2012)\citenamefont
  {Lewenstein}, \citenamefont {Sanpera},\ and\ \citenamefont
  {Ahufinger}}]{Lewenstein}%
  \BibitemOpen
  \bibfield  {author} {\bibinfo {author} {\bibfnamefont {M.}~\bibnamefont
  {Lewenstein}}, \bibinfo {author} {\bibfnamefont {A.}~\bibnamefont {Sanpera}},
  \ and\ \bibinfo {author} {\bibfnamefont {V.}~\bibnamefont {Ahufinger}},\
  }\href@noop {} {\emph {\bibinfo {title} {Ultracold Atoms in Optical
  Lattices}}}\ (\bibinfo  {publisher} {Oxford University Press},\ \bibinfo
  {address} {Oxford, United Kingdom},\ \bibinfo {year} {2012})\BibitemShut
  {NoStop}%
\bibitem [{\citenamefont {Kane}\ and\ \citenamefont
  {Mele}(2005)}]{Kane2005prl}%
  \BibitemOpen
  \bibfield  {author} {\bibinfo {author} {\bibfnamefont {C.~L.}\ \bibnamefont
  {Kane}}\ and\ \bibinfo {author} {\bibfnamefont {E.~J.}\ \bibnamefont
  {Mele}},\ }\href {\doibase 10.1103/PhysRevLett.95.226801} {\bibfield
  {journal} {\bibinfo  {journal} {Phys. Rev. Lett.}\ }\textbf {\bibinfo
  {volume} {95}},\ \bibinfo {pages} {226801} (\bibinfo {year}
  {2005})}\BibitemShut {NoStop}%
\bibitem [{\citenamefont {K\"{o}nig}\ \emph {et~al.}(2007)\citenamefont
  {K\"{o}nig}, \citenamefont {Wiedmann}, \citenamefont {Br\"{u}ne}, \citenamefont
  {Roth}, \citenamefont {Buhmann}, \citenamefont {Molenkamp}, \citenamefont
  {Qi},\ and\ \citenamefont {Zhang}}]{Konig2007}%
  \BibitemOpen
  \bibfield  {author} {\bibinfo {author} {\bibfnamefont {M.}~\bibnamefont
  {K\"{o}nig}}, \bibinfo {author} {\bibfnamefont {S.}~\bibnamefont {Wiedmann}},
  \bibinfo {author} {\bibfnamefont {C.}~\bibnamefont {Br\"{u}ne}}, \bibinfo
  {author} {\bibfnamefont {A.}~\bibnamefont {Roth}}, \bibinfo {author}
  {\bibfnamefont {H.}~\bibnamefont {Buhmann}}, \bibinfo {author} {\bibfnamefont
  {L.~W.}\ \bibnamefont {Molenkamp}}, \bibinfo {author} {\bibfnamefont {X.-L.}\
  \bibnamefont {Qi}}, \ and\ \bibinfo {author} {\bibfnamefont {S.-C.}\
  \bibnamefont {Zhang}},\ }\href {\doibase 10.1126/science.1148047} {\bibfield
  {journal} {\bibinfo  {journal} {Science}\ }\textbf {\bibinfo {volume}
  {318}},\ \bibinfo {pages} {766} (\bibinfo {year} {2007})}\BibitemShut
  {NoStop}%
\bibitem [{\citenamefont {Regnault}\ and\ \citenamefont
  {Bernevig}(2011)}]{Regnault2011prx}%
  \BibitemOpen
  \bibfield  {author} {\bibinfo {author} {\bibfnamefont {N.}~\bibnamefont
  {Regnault}}\ and\ \bibinfo {author} {\bibfnamefont {B.~A.}\ \bibnamefont
  {Bernevig}},\ }\href {\doibase 10.1103/PhysRevX.1.021014} {\bibfield
  {journal} {\bibinfo  {journal} {Phys. Rev. X}\ }\textbf {\bibinfo {volume}
  {1}},\ \bibinfo {pages} {021014} (\bibinfo {year} {2011})}\BibitemShut
  {NoStop}%
\bibitem [{\citenamefont {M\"{o}ller}\ and\ \citenamefont
  {Cooper}(2009)}]{Moller2009prl}%
  \BibitemOpen
  \bibfield  {author} {\bibinfo {author} {\bibfnamefont {G.}~\bibnamefont
  {M\"{o}ller}}\ and\ \bibinfo {author} {\bibfnamefont {N.~R.}\ \bibnamefont
  {Cooper}},\ }\href {\doibase 10.1103/PhysRevLett.103.105303} {\bibfield
  {journal} {\bibinfo  {journal} {Phys. Rev. Lett.}\ }\textbf {\bibinfo
  {volume} {103}},\ \bibinfo {pages} {105303} (\bibinfo {year}
  {2009})}\BibitemShut {NoStop}%
\bibitem [{\citenamefont {Fu}\ and\ \citenamefont {Kane}(2008)}]{Fu2008prl}%
  \BibitemOpen
  \bibfield  {author} {\bibinfo {author} {\bibfnamefont {L.}~\bibnamefont
  {Fu}}\ and\ \bibinfo {author} {\bibfnamefont {C.~L.}\ \bibnamefont {Kane}},\
  }\href {\doibase 10.1103/PhysRevLett.100.096407} {\bibfield  {journal}
  {\bibinfo  {journal} {Phys. Rev. Lett.}\ }\textbf {\bibinfo {volume} {100}},\
  \bibinfo {pages} {096407} (\bibinfo {year} {2008})}\BibitemShut {NoStop}%
\bibitem [{\citenamefont {Mourik}\ \emph {et~al.}(2012)\citenamefont {Mourik},
  \citenamefont {Zuo}, \citenamefont {Frolov}, \citenamefont {Plissard},
  \citenamefont {Bakkers},\ and\ \citenamefont {Kouwenhoven}}]{Mourik2012sci}%
  \BibitemOpen
  \bibfield  {author} {\bibinfo {author} {\bibfnamefont {V.}~\bibnamefont
  {Mourik}}, \bibinfo {author} {\bibfnamefont {K.}~\bibnamefont {Zuo}},
  \bibinfo {author} {\bibfnamefont {S.~M.}\ \bibnamefont {Frolov}}, \bibinfo
  {author} {\bibfnamefont {S.~R.}\ \bibnamefont {Plissard}}, \bibinfo {author}
  {\bibfnamefont {E.~P. A.~M.}\ \bibnamefont {Bakkers}}, \ and\ \bibinfo
  {author} {\bibfnamefont {L.~P.}\ \bibnamefont {Kouwenhoven}},\ }\href
  {\doibase 10.1126/science.1222360} {\bibfield  {journal} {\bibinfo  {journal}
  {Science}\ }\textbf {\bibinfo {volume} {336}},\ \bibinfo {pages} {1003}
  (\bibinfo {year} {2012})}\BibitemShut {NoStop}%
\bibitem [{\citenamefont {Hofstadter}(1976)}]{Hofstadter:1976ys}%
  \BibitemOpen
  \bibfield  {author} {\bibinfo {author} {\bibfnamefont {D.~R.}\ \bibnamefont
  {Hofstadter}},\ }\href {\doibase 10.1103/PhysRevB.14.2239} {\bibfield
  {journal} {\bibinfo  {journal} {Phys. Rev. B}\ }\textbf {\bibinfo {volume}
  {14}},\ \bibinfo {pages} {2239} (\bibinfo {year} {1976})}\BibitemShut
  {NoStop}%
\bibitem [{\citenamefont {Hunt}\ \emph {et~al.}(2013)\citenamefont {Hunt},
  \citenamefont {Sanchez-Yamagishi}, \citenamefont {Young}, \citenamefont
  {Yankowitz}, \citenamefont {LeRoy}, \citenamefont {Watanabe}, \citenamefont
  {Taniguchi}, \citenamefont {Moon}, \citenamefont {Koshino}, \citenamefont
  {Jarillo-Herrero},\ and\ \citenamefont {Ashoori}}]{Hunt2013}%
  \BibitemOpen
  \bibfield  {author} {\bibinfo {author} {\bibfnamefont {B.}~\bibnamefont
  {Hunt}}, \bibinfo {author} {\bibfnamefont {J.~D.}\ \bibnamefont
  {Sanchez-Yamagishi}}, \bibinfo {author} {\bibfnamefont {A.~F.}\ \bibnamefont
  {Young}}, \bibinfo {author} {\bibfnamefont {M.}~\bibnamefont {Yankowitz}},
  \bibinfo {author} {\bibfnamefont {B.~J.}\ \bibnamefont {LeRoy}}, \bibinfo
  {author} {\bibfnamefont {K.}~\bibnamefont {Watanabe}}, \bibinfo {author}
  {\bibfnamefont {T.}~\bibnamefont {Taniguchi}}, \bibinfo {author}
  {\bibfnamefont {P.}~\bibnamefont {Moon}}, \bibinfo {author} {\bibfnamefont
  {M.}~\bibnamefont {Koshino}}, \bibinfo {author} {\bibfnamefont
  {P.}~\bibnamefont {Jarillo-Herrero}}, \ and\ \bibinfo {author} {\bibfnamefont
  {R.~C.}\ \bibnamefont {Ashoori}},\ }\href {\doibase 10.1126/science.1237240}
  {\bibfield  {journal} {\bibinfo  {journal} {Science}\ }\textbf {\bibinfo
  {volume} {340}},\ \bibinfo {pages} {1427} (\bibinfo {year}
  {2013})}\BibitemShut {NoStop}%
\bibitem [{\citenamefont {Ponomarenko}\ \emph {et~al.}(2013)\citenamefont
  {Ponomarenko}, \citenamefont {Gorbachev}, \citenamefont {Yu}, \citenamefont
  {Elias}, \citenamefont {Jalil}, \citenamefont {Patel}, \citenamefont
  {Mishchenko}, \citenamefont {Mayorov}, \citenamefont {Woods}, \citenamefont
  {Wallbank}, \citenamefont {Mucha-Kruczynski}, \citenamefont {Piot},
  \citenamefont {Potemski}, \citenamefont {Grigorieva}, \citenamefont
  {Novoselov}, \citenamefont {Guinea}, \citenamefont {Falko},\ and\
  \citenamefont {Geim}}]{Novoselov2013}%
  \BibitemOpen
  \bibfield  {author} {\bibinfo {author} {\bibfnamefont {L.}~\bibnamefont
  {Ponomarenko}}, \bibinfo {author} {\bibfnamefont {R.~V.}\ \bibnamefont
  {Gorbachev}}, \bibinfo {author} {\bibfnamefont {G.~L.}\ \bibnamefont {Yu}},
  \bibinfo {author} {\bibfnamefont {D.~C.}\ \bibnamefont {Elias}}, \bibinfo
  {author} {\bibfnamefont {R.}~\bibnamefont {Jalil}}, \bibinfo {author}
  {\bibfnamefont {A.~A.}\ \bibnamefont {Patel}}, \bibinfo {author}
  {\bibfnamefont {A.}~\bibnamefont {Mishchenko}}, \bibinfo {author}
  {\bibfnamefont {A.~S.}\ \bibnamefont {Mayorov}}, \bibinfo {author}
  {\bibfnamefont {C.~R.}\ \bibnamefont {Woods}}, \bibinfo {author}
  {\bibfnamefont {J.~R.}\ \bibnamefont {Wallbank}}, \bibinfo {author}
  {\bibfnamefont {M.}~\bibnamefont {Mucha-Kruczynski}}, \bibinfo {author}
  {\bibfnamefont {B.~A.}\ \bibnamefont {Piot}}, \bibinfo {author}
  {\bibfnamefont {M.}~\bibnamefont {Potemski}}, \bibinfo {author}
  {\bibfnamefont {I.}~\bibnamefont {Grigorieva}}, \bibinfo {author}
  {\bibfnamefont {K.~S.}\ \bibnamefont {Novoselov}}, \bibinfo {author}
  {\bibfnamefont {F.}~\bibnamefont {Guinea}}, \bibinfo {author} {\bibfnamefont
  {V.~I.}\ \bibnamefont {Falko}}, \ and\ \bibinfo {author} {\bibfnamefont
  {A.~K.}\ \bibnamefont {Geim}},\ }\href
  {http://dx.doi.org/10.1038/nature12187} {\bibfield  {journal} {\bibinfo
  {journal} {Nature}\ }\textbf {\bibinfo {volume} {497}},\ \bibinfo {pages}
  {594} (\bibinfo {year} {2013})}\BibitemShut {NoStop}%
\bibitem [{\citenamefont {Dean}\ \emph {et~al.}(2013)\citenamefont {Dean},
  \citenamefont {Wang}, \citenamefont {Maher}, \citenamefont {Forsythe},
  \citenamefont {Ghahari}, \citenamefont {Gao}, \citenamefont {Katoch},
  \citenamefont {Ishigami}, \citenamefont {Moon}, \citenamefont {Koshino},
  \citenamefont {Taniguchi}, \citenamefont {Watanabe}, \citenamefont {Shepard},
  \citenamefont {Hone},\ and\ \citenamefont {Kim}}]{Dean2013nature}%
  \BibitemOpen
  \bibfield  {author} {\bibinfo {author} {\bibfnamefont {C.~R.}\ \bibnamefont
  {Dean}}, \bibinfo {author} {\bibfnamefont {L.}~\bibnamefont {Wang}}, \bibinfo
  {author} {\bibfnamefont {P.}~\bibnamefont {Maher}}, \bibinfo {author}
  {\bibfnamefont {C.}~\bibnamefont {Forsythe}}, \bibinfo {author}
  {\bibfnamefont {F.}~\bibnamefont {Ghahari}}, \bibinfo {author} {\bibfnamefont
  {Y.}~\bibnamefont {Gao}}, \bibinfo {author} {\bibfnamefont {J.}~\bibnamefont
  {Katoch}}, \bibinfo {author} {\bibfnamefont {M.}~\bibnamefont {Ishigami}},
  \bibinfo {author} {\bibfnamefont {P.}~\bibnamefont {Moon}}, \bibinfo {author}
  {\bibfnamefont {M.}~\bibnamefont {Koshino}}, \bibinfo {author} {\bibfnamefont
  {T.}~\bibnamefont {Taniguchi}}, \bibinfo {author} {\bibfnamefont
  {K.}~\bibnamefont {Watanabe}}, \bibinfo {author} {\bibfnamefont {K.~L.}\
  \bibnamefont {Shepard}}, \bibinfo {author} {\bibfnamefont {J.}~\bibnamefont
  {Hone}}, \ and\ \bibinfo {author} {\bibfnamefont {P.}~\bibnamefont {Kim}},\
  }\href {http://www.nature.com/nature/journal/v497/n7451/abs/nature12186.html}
  {\bibfield  {journal} {\bibinfo  {journal} {Nature}\ }\textbf {\bibinfo
  {volume} {497}},\ \bibinfo {pages} {598} (\bibinfo {year}
  {2013})}\BibitemShut {NoStop}%
\bibitem [{\citenamefont {Albrecht}\ \emph {et~al.}(2001)\citenamefont
  {Albrecht}, \citenamefont {Smet}, \citenamefont {von Klitzing}, \citenamefont
  {Weiss}, \citenamefont {Umansky},\ and\ \citenamefont
  {Schweizer}}]{vonKlitzing2001prl}%
  \BibitemOpen
  \bibfield  {author} {\bibinfo {author} {\bibfnamefont {C.}~\bibnamefont
  {Albrecht}}, \bibinfo {author} {\bibfnamefont {J.~H.}\ \bibnamefont {Smet}},
  \bibinfo {author} {\bibfnamefont {K.}~\bibnamefont {von Klitzing}}, \bibinfo
  {author} {\bibfnamefont {D.}~\bibnamefont {Weiss}}, \bibinfo {author}
  {\bibfnamefont {V.}~\bibnamefont {Umansky}}, \ and\ \bibinfo {author}
  {\bibfnamefont {H.}~\bibnamefont {Schweizer}},\ }\href {\doibase
  10.1103/PhysRevLett.86.147} {\bibfield  {journal} {\bibinfo  {journal} {Phys.
  Rev. Lett.}\ }\textbf {\bibinfo {volume} {86}},\ \bibinfo {pages} {147}
  (\bibinfo {year} {2001})}\BibitemShut {NoStop}%
\bibitem [{\citenamefont {Madison}\ \emph {et~al.}(2000)\citenamefont
  {Madison}, \citenamefont {Chevy}, \citenamefont {Wohlleben},\ and\
  \citenamefont {Dalibard}}]{Dalibard2000}%
  \BibitemOpen
  \bibfield  {author} {\bibinfo {author} {\bibfnamefont {K.~W.}\ \bibnamefont
  {Madison}}, \bibinfo {author} {\bibfnamefont {F.}~\bibnamefont {Chevy}},
  \bibinfo {author} {\bibfnamefont {W.}~\bibnamefont {Wohlleben}}, \ and\
  \bibinfo {author} {\bibfnamefont {J.}~\bibnamefont {Dalibard}},\ }\href
  {\doibase 10.1103/PhysRevLett.84.806} {\bibfield  {journal} {\bibinfo
  {journal} {Phys. Rev. Lett.}\ }\textbf {\bibinfo {volume} {84}},\ \bibinfo
  {pages} {806} (\bibinfo {year} {2000})}\BibitemShut {NoStop}%
\bibitem [{\citenamefont {Abo-Shaeer}\ \emph {et~al.}(2001)\citenamefont
  {Abo-Shaeer}, \citenamefont {Raman}, \citenamefont {Vogels},\ and\
  \citenamefont {Ketterle}}]{AboShaeer2001}%
  \BibitemOpen
  \bibfield  {author} {\bibinfo {author} {\bibfnamefont {J.~R.}\ \bibnamefont
  {Abo-Shaeer}}, \bibinfo {author} {\bibfnamefont {C.}~\bibnamefont {Raman}},
  \bibinfo {author} {\bibfnamefont {J.~M.}\ \bibnamefont {Vogels}}, \ and\
  \bibinfo {author} {\bibfnamefont {W.}~\bibnamefont {Ketterle}},\ }\href
  {\doibase 10.1126/science.1060182} {\bibfield  {journal} {\bibinfo  {journal}
  {Science}\ }\textbf {\bibinfo {volume} {292}},\ \bibinfo {pages} {476}
  (\bibinfo {year} {2001})}\BibitemShut {NoStop}%
\bibitem [{\citenamefont {Lin}\ \emph {et~al.}(2009)\citenamefont {Lin},
  \citenamefont {Compton}, \citenamefont {Jimenez-Garcia}, \citenamefont
  {Porto},\ and\ \citenamefont {Spielman}}]{Lin2009}%
  \BibitemOpen
  \bibfield  {author} {\bibinfo {author} {\bibfnamefont {Y.-J.}\ \bibnamefont
  {Lin}}, \bibinfo {author} {\bibfnamefont {R.~L.}\ \bibnamefont {Compton}},
  \bibinfo {author} {\bibfnamefont {K.}~\bibnamefont {Jimenez-Garcia}},
  \bibinfo {author} {\bibfnamefont {J.~V.}\ \bibnamefont {Porto}}, \ and\
  \bibinfo {author} {\bibfnamefont {I.}~\bibnamefont {Spielman}},\ }\href
  {http://dx.doi.org/10.1038/nature08609} {\bibfield  {journal} {\bibinfo
  {journal} {Nature}\ }\textbf {\bibinfo {volume} {462}},\ \bibinfo {pages}
  {628} (\bibinfo {year} {2009})}\BibitemShut {NoStop}%
\bibitem [{\citenamefont {Dalibard}\ \emph {et~al.}(2011)\citenamefont
  {Dalibard}, \citenamefont {Gerbier}, \citenamefont
  {Juzeli\ifmmode~\bar{u}\else \={u}\fi{}nas},\ and\ \citenamefont
  {\"Ohberg}}]{Dalibard2011rmp}%
  \BibitemOpen
  \bibfield  {author} {\bibinfo {author} {\bibfnamefont {J.}~\bibnamefont
  {Dalibard}}, \bibinfo {author} {\bibfnamefont {F.}~\bibnamefont {Gerbier}},
  \bibinfo {author} {\bibfnamefont {G.}~\bibnamefont
  {Juzeli\ifmmode~\bar{u}\else \={u}\fi{}nas}}, \ and\ \bibinfo {author}
  {\bibfnamefont {P.}~\bibnamefont {\"Ohberg}},\ }\href {\doibase
  10.1103/RevModPhys.83.1523} {\bibfield  {journal} {\bibinfo  {journal} {Rev.
  Mod. Phys.}\ }\textbf {\bibinfo {volume} {83}},\ \bibinfo {pages} {1523}
  (\bibinfo {year} {2011})}\BibitemShut {NoStop}%
\bibitem [{\citenamefont {Jaksch}\ and\ \citenamefont
  {Zoller}(2003)}]{Jaksch:2003fk}%
  \BibitemOpen
  \bibfield  {author} {\bibinfo {author} {\bibfnamefont {D.}~\bibnamefont
  {Jaksch}}\ and\ \bibinfo {author} {\bibfnamefont {P.}~\bibnamefont
  {Zoller}},\ }\href {http://stacks.iop.org/1367-2630/5/i=1/a=356} {\bibfield
  {journal} {\bibinfo  {journal} {New Journal of Physics}\ }\textbf {\bibinfo
  {volume} {5}},\ \bibinfo {pages} {56} (\bibinfo {year} {2003})}\BibitemShut
  {NoStop}%
\bibitem [{\citenamefont {Mueller}(2004)}]{Mueller2004pra}%
  \BibitemOpen
  \bibfield  {author} {\bibinfo {author} {\bibfnamefont {E.~J.}\ \bibnamefont
  {Mueller}},\ }\href {\doibase 10.1103/PhysRevA.70.041603} {\bibfield
  {journal} {\bibinfo  {journal} {Phys. Rev. A}\ }\textbf {\bibinfo {volume}
  {70}},\ \bibinfo {pages} {041603} (\bibinfo {year} {2004})}\BibitemShut
  {NoStop}%
\bibitem [{\citenamefont {Gerbier}\ and\ \citenamefont
  {Dalibard}(2010)}]{Gerbier2010njp}%
  \BibitemOpen
  \bibfield  {author} {\bibinfo {author} {\bibfnamefont {F.}~\bibnamefont
  {Gerbier}}\ and\ \bibinfo {author} {\bibfnamefont {J.}~\bibnamefont
  {Dalibard}},\ }\href {http://stacks.iop.org/1367-2630/12/i=3/a=033007}
  {\bibfield  {journal} {\bibinfo  {journal} {New Journal of Physics}\ }\textbf
  {\bibinfo {volume} {12}},\ \bibinfo {pages} {033007} (\bibinfo {year}
  {2010})}\BibitemShut {NoStop}%
\bibitem [{\citenamefont {Struck}\ \emph {et~al.}(2012)\citenamefont {Struck},
  \citenamefont {\"Olschl\"ager}, \citenamefont {Weinberg}, \citenamefont
  {Hauke}, \citenamefont {Simonet}, \citenamefont {Eckardt}, \citenamefont
  {Lewenstein}, \citenamefont {Sengstock},\ and\ \citenamefont
  {Windpassinger}}]{Struck2012}%
  \BibitemOpen
  \bibfield  {author} {\bibinfo {author} {\bibfnamefont {J.}~\bibnamefont
  {Struck}}, \bibinfo {author} {\bibfnamefont {C.}~\bibnamefont
  {\"Olschl\"ager}}, \bibinfo {author} {\bibfnamefont {M.}~\bibnamefont
  {Weinberg}}, \bibinfo {author} {\bibfnamefont {P.}~\bibnamefont {Hauke}},
  \bibinfo {author} {\bibfnamefont {J.}~\bibnamefont {Simonet}}, \bibinfo
  {author} {\bibfnamefont {A.}~\bibnamefont {Eckardt}}, \bibinfo {author}
  {\bibfnamefont {M.}~\bibnamefont {Lewenstein}}, \bibinfo {author}
  {\bibfnamefont {K.}~\bibnamefont {Sengstock}}, \ and\ \bibinfo {author}
  {\bibfnamefont {P.}~\bibnamefont {Windpassinger}},\ }\href {\doibase
  10.1103/PhysRevLett.108.225304} {\bibfield  {journal} {\bibinfo  {journal}
  {Phys. Rev. Lett.}\ }\textbf {\bibinfo {volume} {108}},\ \bibinfo {pages}
  {225304} (\bibinfo {year} {2012})}\BibitemShut {NoStop}%
\bibitem [{\citenamefont {{Struck}}\ \emph {et~al.}(2013)\citenamefont
  {{Struck}}, \citenamefont {{Weinberg}}, \citenamefont {{{\"O}lschl{\"a}ger}},
  \citenamefont {{Windpassinger}}, \citenamefont {{Simonet}}, \citenamefont
  {{Sengstock}}, \citenamefont {{H{\"o}ppner}}, \citenamefont {{Hauke}},
  \citenamefont {{Eckardt}}, \citenamefont {{Lewenstein}},\ and\ \citenamefont
  {{Mathey}}}]{Struck:2013ly}%
  \BibitemOpen
  \bibfield  {author} {\bibinfo {author} {\bibfnamefont {J.}~\bibnamefont
  {{Struck}}}, \bibinfo {author} {\bibfnamefont {M.}~\bibnamefont
  {{Weinberg}}}, \bibinfo {author} {\bibfnamefont {C.}~\bibnamefont
  {{{\"O}lschl{\"a}ger}}}, \bibinfo {author} {\bibfnamefont {P.}~\bibnamefont
  {{Windpassinger}}}, \bibinfo {author} {\bibfnamefont {J.}~\bibnamefont
  {{Simonet}}}, \bibinfo {author} {\bibfnamefont {K.}~\bibnamefont
  {{Sengstock}}}, \bibinfo {author} {\bibfnamefont {R.}~\bibnamefont
  {{H{\"o}ppner}}}, \bibinfo {author} {\bibfnamefont {P.}~\bibnamefont
  {{Hauke}}}, \bibinfo {author} {\bibfnamefont {A.}~\bibnamefont {{Eckardt}}},
  \bibinfo {author} {\bibfnamefont {M.}~\bibnamefont {{Lewenstein}}}, \ and\
  \bibinfo {author} {\bibfnamefont {L.}~\bibnamefont {{Mathey}}},\ }\href@noop
  {} {\bibfield  {journal} {\bibinfo  {journal} {ArXiv e-prints}\ } (\bibinfo
  {year} {2013})},\ \Eprint {http://arxiv.org/abs/1304.5520} {arXiv:1304.5520
  [cond-mat.quant-gas]} \BibitemShut {NoStop}%
\bibitem [{\citenamefont {Aidelsburger}\ \emph {et~al.}(2011)\citenamefont
  {Aidelsburger}, \citenamefont {Atala}, \citenamefont {Nascimb\`ene},
  \citenamefont {Trotzky}, \citenamefont {Chen},\ and\ \citenamefont
  {Bloch}}]{Aidelsburger2011}%
  \BibitemOpen
  \bibfield  {author} {\bibinfo {author} {\bibfnamefont {M.}~\bibnamefont
  {Aidelsburger}}, \bibinfo {author} {\bibfnamefont {M.}~\bibnamefont {Atala}},
  \bibinfo {author} {\bibfnamefont {S.}~\bibnamefont {Nascimb\`ene}}, \bibinfo
  {author} {\bibfnamefont {S.}~\bibnamefont {Trotzky}}, \bibinfo {author}
  {\bibfnamefont {Y.-A.}\ \bibnamefont {Chen}}, \ and\ \bibinfo {author}
  {\bibfnamefont {I.}~\bibnamefont {Bloch}},\ }\href {\doibase
  10.1103/PhysRevLett.107.255301} {\bibfield  {journal} {\bibinfo  {journal}
  {Phys. Rev. Lett.}\ }\textbf {\bibinfo {volume} {107}},\ \bibinfo {pages}
  {255301} (\bibinfo {year} {2011})}\BibitemShut {NoStop}%
\bibitem [{\citenamefont {Harper}(1955)}]{Harper1955}%
  \BibitemOpen
  \bibfield  {author} {\bibinfo {author} {\bibfnamefont {P.~G.}\ \bibnamefont
  {Harper}},\ }\href {http://stacks.iop.org/0370-1298/68/i=10/a=304} {\bibfield
   {journal} {\bibinfo  {journal} {Proceedings of the Physical Society. Section
  A}\ }\textbf {\bibinfo {volume} {68}},\ \bibinfo {pages} {874} (\bibinfo
  {year} {1955})}\BibitemShut {NoStop}%
\bibitem [{\citenamefont {Kolovsky}(2011)}]{Kolovsky:2011bh}%
  \BibitemOpen
  \bibfield  {author} {\bibinfo {author} {\bibfnamefont {A.~R.}\ \bibnamefont
  {Kolovsky}},\ }\href {http://stacks.iop.org/0295-5075/93/i=2/a=20003}
  {\bibfield  {journal} {\bibinfo  {journal} {Europhysics Letters}\ }\textbf
  {\bibinfo {volume} {93}},\ \bibinfo {pages} {20003} (\bibinfo {year}
  {2011})}\BibitemShut {NoStop}%
\bibitem [{\citenamefont {Creffield}\ and\ \citenamefont
  {Sols}(2013)}]{creffield2013epl}%
  \BibitemOpen
  \bibfield  {author} {\bibinfo {author} {\bibfnamefont {C.~E.}\ \bibnamefont
  {Creffield}}\ and\ \bibinfo {author} {\bibfnamefont {F.}~\bibnamefont
  {Sols}},\ }\href {http://stacks.iop.org/0295-5075/101/i=4/a=40001} {\bibfield
   {journal} {\bibinfo  {journal} {Europhysics Letters}\ }\textbf {\bibinfo
  {volume} {101}},\ \bibinfo {pages} {40001} (\bibinfo {year}
  {2013})}\BibitemShut {NoStop}%
\bibitem [{\citenamefont {Miyake}(2013)}]{Miyake2013}%
  \BibitemOpen
  \bibfield  {author} {\bibinfo {author} {\bibfnamefont {H.}~\bibnamefont
  {Miyake}},\ }\emph {\bibinfo {title} {{Probing and Preparing Novel States of
  Quantum Degenerate Rubidium Atoms in Optical Lattices}}},\ \href@noop {}
  {Ph.D. thesis},\ \bibinfo  {school} {Massachusetts Institute of Technology}
  (\bibinfo {year} {2013})\BibitemShut {NoStop}%
\bibitem [{\citenamefont {Aidelsburger}\ \emph {et~al.}(2013)\citenamefont
  {Aidelsburger}, \citenamefont {Atala}, \citenamefont {Nascimb{\`e}ne},
  \citenamefont {Trotzky}, \citenamefont {Chen},\ and\ \citenamefont
  {Bloch}}]{Aidelsburger:2013ys}%
  \BibitemOpen
  \bibfield  {author} {\bibinfo {author} {\bibfnamefont {M.}~\bibnamefont
  {Aidelsburger}}, \bibinfo {author} {\bibfnamefont {M.}~\bibnamefont {Atala}},
  \bibinfo {author} {\bibfnamefont {S.}~\bibnamefont {Nascimb{\`e}ne}},
  \bibinfo {author} {\bibfnamefont {S.}~\bibnamefont {Trotzky}}, \bibinfo
  {author} {\bibfnamefont {Y.-A.}\ \bibnamefont {Chen}}, \ and\ \bibinfo
  {author} {\bibfnamefont {I.}~\bibnamefont {Bloch}},\ }\href {\doibase
  10.1007/s00340-013-5418-1} {\bibfield  {journal} {\bibinfo  {journal}
  {Applied Physics B}\ } (\bibinfo {year} {2013}),\
  10.1007/s00340-013-5418-1}\BibitemShut {NoStop}%
\bibitem [{\citenamefont {Bermudez}\ \emph {et~al.}(2011)\citenamefont
  {Bermudez}, \citenamefont {Schaetz},\ and\ \citenamefont
  {Porras}}]{Bermudez2011prl}%
  \BibitemOpen
  \bibfield  {author} {\bibinfo {author} {\bibfnamefont {A.}~\bibnamefont
  {Bermudez}}, \bibinfo {author} {\bibfnamefont {T.}~\bibnamefont {Schaetz}}, \
  and\ \bibinfo {author} {\bibfnamefont {D.}~\bibnamefont {Porras}},\ }\href
  {\doibase 10.1103/PhysRevLett.107.150501} {\bibfield  {journal} {\bibinfo
  {journal} {Phys. Rev. Lett.}\ }\textbf {\bibinfo {volume} {107}},\ \bibinfo
  {pages} {150501} (\bibinfo {year} {2011})}\BibitemShut {NoStop}%
\bibitem [{\citenamefont {Tarallo}\ \emph {et~al.}(2012)\citenamefont
  {Tarallo}, \citenamefont {Alberti}, \citenamefont {Poli}, \citenamefont
  {Chiofalo}, \citenamefont {Wang},\ and\ \citenamefont {Tino}}]{Tino2012}%
  \BibitemOpen
  \bibfield  {author} {\bibinfo {author} {\bibfnamefont {M.~G.}\ \bibnamefont
  {Tarallo}}, \bibinfo {author} {\bibfnamefont {A.}~\bibnamefont {Alberti}},
  \bibinfo {author} {\bibfnamefont {N.}~\bibnamefont {Poli}}, \bibinfo {author}
  {\bibfnamefont {M.~L.}\ \bibnamefont {Chiofalo}}, \bibinfo {author}
  {\bibfnamefont {F.-Y.}\ \bibnamefont {Wang}}, \ and\ \bibinfo {author}
  {\bibfnamefont {G.~M.}\ \bibnamefont {Tino}},\ }\href {\doibase
  10.1103/PhysRevA.86.033615} {\bibfield  {journal} {\bibinfo  {journal} {Phys.
  Rev. A}\ }\textbf {\bibinfo {volume} {86}},\ \bibinfo {pages} {033615}
  (\bibinfo {year} {2012})}\BibitemShut {NoStop}%
\bibitem [{\citenamefont {{Aidelsburger}}\ \emph {et~al.}(2013)\citenamefont
  {{Aidelsburger}}, \citenamefont {{Atala}}, \citenamefont {{Lohse}},
  \citenamefont {{Barreiro}}, \citenamefont {{Paredes}},\ and\ \citenamefont
  {{Bloch}}}]{Aidelsburger2013arxiv}%
  \BibitemOpen
  \bibfield  {author} {\bibinfo {author} {\bibfnamefont {M.}~\bibnamefont
  {{Aidelsburger}}}, \bibinfo {author} {\bibfnamefont {M.}~\bibnamefont
  {{Atala}}}, \bibinfo {author} {\bibfnamefont {M.}~\bibnamefont {{Lohse}}},
  \bibinfo {author} {\bibfnamefont {J.~T.}\ \bibnamefont {{Barreiro}}},
  \bibinfo {author} {\bibfnamefont {B.}~\bibnamefont {{Paredes}}}, \ and\
  \bibinfo {author} {\bibfnamefont {I.}~\bibnamefont {{Bloch}}},\ }\href@noop
  {} {\bibfield  {journal} {\bibinfo  {journal} {ArXiv e-prints}\ } (\bibinfo
  {year} {2013})},\ \Eprint {http://arxiv.org/abs/1308.0321} {arXiv:1308.0321
  [cond-mat.quant-gas]} \BibitemShut {NoStop}%
\bibitem [{\citenamefont {Sias}\ \emph {et~al.}(2008)\citenamefont {Sias},
  \citenamefont {Lignier}, \citenamefont {Singh}, \citenamefont {Zenesini},
  \citenamefont {Ciampini}, \citenamefont {Morsch},\ and\ \citenamefont
  {Arimondo}}]{Sias2008}%
  \BibitemOpen
  \bibfield  {author} {\bibinfo {author} {\bibfnamefont {C.}~\bibnamefont
  {Sias}}, \bibinfo {author} {\bibfnamefont {H.}~\bibnamefont {Lignier}},
  \bibinfo {author} {\bibfnamefont {Y.~P.}\ \bibnamefont {Singh}}, \bibinfo
  {author} {\bibfnamefont {A.}~\bibnamefont {Zenesini}}, \bibinfo {author}
  {\bibfnamefont {D.}~\bibnamefont {Ciampini}}, \bibinfo {author}
  {\bibfnamefont {O.}~\bibnamefont {Morsch}}, \ and\ \bibinfo {author}
  {\bibfnamefont {E.}~\bibnamefont {Arimondo}},\ }\href {\doibase
  10.1103/PhysRevLett.100.040404} {\bibfield  {journal} {\bibinfo  {journal}
  {Phys. Rev. Lett.}\ }\textbf {\bibinfo {volume} {100}},\ \bibinfo {pages}
  {040404} (\bibinfo {year} {2008})}\BibitemShut {NoStop}%
\bibitem [{\citenamefont {Ivanov}\ \emph {et~al.}(2008)\citenamefont {Ivanov},
  \citenamefont {Alberti}, \citenamefont {Schioppo}, \citenamefont {Ferrari},
  \citenamefont {Artoni}, \citenamefont {Chiofalo},\ and\ \citenamefont
  {Tino}}]{Ivanov2008}%
  \BibitemOpen
  \bibfield  {author} {\bibinfo {author} {\bibfnamefont {V.~V.}\ \bibnamefont
  {Ivanov}}, \bibinfo {author} {\bibfnamefont {A.}~\bibnamefont {Alberti}},
  \bibinfo {author} {\bibfnamefont {M.}~\bibnamefont {Schioppo}}, \bibinfo
  {author} {\bibfnamefont {G.}~\bibnamefont {Ferrari}}, \bibinfo {author}
  {\bibfnamefont {M.}~\bibnamefont {Artoni}}, \bibinfo {author} {\bibfnamefont
  {M.~L.}\ \bibnamefont {Chiofalo}}, \ and\ \bibinfo {author} {\bibfnamefont
  {G.~M.}\ \bibnamefont {Tino}},\ }\href {\doibase
  10.1103/PhysRevLett.100.043602} {\bibfield  {journal} {\bibinfo  {journal}
  {Phys. Rev. Lett.}\ }\textbf {\bibinfo {volume} {100}},\ \bibinfo {pages}
  {043602} (\bibinfo {year} {2008})}\BibitemShut {NoStop}%
\bibitem [{\citenamefont {M\"{o}ller}\ and\ \citenamefont
  {Cooper}(2010)}]{Moller2010pra}%
  \BibitemOpen
  \bibfield  {author} {\bibinfo {author} {\bibfnamefont {G.}~\bibnamefont
  {M\"{o}ller}}\ and\ \bibinfo {author} {\bibfnamefont {N.~R.}\ \bibnamefont
  {Cooper}},\ }\href {\doibase 10.1103/PhysRevA.82.063625} {\bibfield
  {journal} {\bibinfo  {journal} {Phys. Rev. A}\ }\textbf {\bibinfo {volume}
  {82}},\ \bibinfo {pages} {063625} (\bibinfo {year} {2010})}\BibitemShut
  {NoStop}%
\bibitem [{\citenamefont {Polak}\ and\ \citenamefont
  {Zaleski}(2013)}]{Polak2013}%
  \BibitemOpen
  \bibfield  {author} {\bibinfo {author} {\bibfnamefont {T.~P.}\ \bibnamefont
  {Polak}}\ and\ \bibinfo {author} {\bibfnamefont {T.~A.}\ \bibnamefont
  {Zaleski}},\ }\href {\doibase 10.1103/PhysRevA.87.033614} {\bibfield
  {journal} {\bibinfo  {journal} {Phys. Rev. A}\ }\textbf {\bibinfo {volume}
  {87}},\ \bibinfo {pages} {033614} (\bibinfo {year} {2013})}\BibitemShut
  {NoStop}%
\bibitem [{\citenamefont {Powell}\ \emph
  {et~al.}(2011{\natexlab{a}})\citenamefont {Powell}, \citenamefont {Barnett},
  \citenamefont {Sensarma},\ and\ \citenamefont {Das~Sarma}}]{Powell2011pra}%
  \BibitemOpen
  \bibfield  {author} {\bibinfo {author} {\bibfnamefont {S.}~\bibnamefont
  {Powell}}, \bibinfo {author} {\bibfnamefont {R.}~\bibnamefont {Barnett}},
  \bibinfo {author} {\bibfnamefont {R.}~\bibnamefont {Sensarma}}, \ and\
  \bibinfo {author} {\bibfnamefont {S.}~\bibnamefont {Das~Sarma}},\ }\href
  {\doibase 10.1103/PhysRevA.83.013612} {\bibfield  {journal} {\bibinfo
  {journal} {Phys. Rev. A}\ }\textbf {\bibinfo {volume} {83}},\ \bibinfo
  {pages} {013612} (\bibinfo {year} {2011}{\natexlab{a}})}\BibitemShut
  {NoStop}%
\bibitem [{\citenamefont {Fallani}\ \emph {et~al.}(2004)\citenamefont
  {Fallani}, \citenamefont {De~Sarlo}, \citenamefont {Lye}, \citenamefont
  {Modugno}, \citenamefont {Saers}, \citenamefont {Fort},\ and\ \citenamefont
  {Inguscio}}]{Fallani2004}%
  \BibitemOpen
  \bibfield  {author} {\bibinfo {author} {\bibfnamefont {L.}~\bibnamefont
  {Fallani}}, \bibinfo {author} {\bibfnamefont {L.}~\bibnamefont {De~Sarlo}},
  \bibinfo {author} {\bibfnamefont {J.~E.}\ \bibnamefont {Lye}}, \bibinfo
  {author} {\bibfnamefont {M.}~\bibnamefont {Modugno}}, \bibinfo {author}
  {\bibfnamefont {R.}~\bibnamefont {Saers}}, \bibinfo {author} {\bibfnamefont
  {C.}~\bibnamefont {Fort}}, \ and\ \bibinfo {author} {\bibfnamefont
  {M.}~\bibnamefont {Inguscio}},\ }\href {\doibase
  10.1103/PhysRevLett.93.140406} {\bibfield  {journal} {\bibinfo  {journal}
  {Phys. Rev. Lett.}\ }\textbf {\bibinfo {volume} {93}},\ \bibinfo {pages}
  {140406} (\bibinfo {year} {2004})}\BibitemShut {NoStop}%
\bibitem [{\citenamefont {Campbell}\ \emph {et~al.}(2006)\citenamefont
  {Campbell}, \citenamefont {Mun}, \citenamefont {Boyd}, \citenamefont
  {Streed}, \citenamefont {Ketterle},\ and\ \citenamefont
  {Pritchard}}]{Campbell2006}%
  \BibitemOpen
  \bibfield  {author} {\bibinfo {author} {\bibfnamefont {G.~K.}\ \bibnamefont
  {Campbell}}, \bibinfo {author} {\bibfnamefont {J.}~\bibnamefont {Mun}},
  \bibinfo {author} {\bibfnamefont {M.}~\bibnamefont {Boyd}}, \bibinfo {author}
  {\bibfnamefont {E.~W.}\ \bibnamefont {Streed}}, \bibinfo {author}
  {\bibfnamefont {W.}~\bibnamefont {Ketterle}}, \ and\ \bibinfo {author}
  {\bibfnamefont {D.~E.}\ \bibnamefont {Pritchard}},\ }\href {\doibase
  10.1103/PhysRevLett.96.020406} {\bibfield  {journal} {\bibinfo  {journal}
  {Phys. Rev. Lett.}\ }\textbf {\bibinfo {volume} {96}},\ \bibinfo {pages}
  {020406} (\bibinfo {year} {2006})}\BibitemShut {NoStop}%
\bibitem [{\citenamefont {Powell}\ \emph
  {et~al.}(2011{\natexlab{b}})\citenamefont {Powell}, \citenamefont {Barnett},
  \citenamefont {Sensarma},\ and\ \citenamefont {Das~Sarma}}]{DasSarma2011}%
  \BibitemOpen
  \bibfield  {author} {\bibinfo {author} {\bibfnamefont {S.}~\bibnamefont
  {Powell}}, \bibinfo {author} {\bibfnamefont {R.}~\bibnamefont {Barnett}},
  \bibinfo {author} {\bibfnamefont {R.}~\bibnamefont {Sensarma}}, \ and\
  \bibinfo {author} {\bibfnamefont {S.}~\bibnamefont {Das~Sarma}},\ }\href
  {\doibase 10.1103/PhysRevA.83.013612} {\bibfield  {journal} {\bibinfo
  {journal} {Phys. Rev. A}\ }\textbf {\bibinfo {volume} {83}},\ \bibinfo
  {pages} {013612} (\bibinfo {year} {2011}{\natexlab{b}})}\BibitemShut
  {NoStop}%
\bibitem [{\citenamefont {S\o{}rensen}\ \emph {et~al.}(2005)\citenamefont
  {S\o{}rensen}, \citenamefont {Demler},\ and\ \citenamefont
  {Lukin}}]{Lukin2005prl}%
  \BibitemOpen
  \bibfield  {author} {\bibinfo {author} {\bibfnamefont {A.~S.}\ \bibnamefont
  {S\o{}rensen}}, \bibinfo {author} {\bibfnamefont {E.}~\bibnamefont {Demler}},
  \ and\ \bibinfo {author} {\bibfnamefont {M.~D.}\ \bibnamefont {Lukin}},\
  }\href {\doibase 10.1103/PhysRevLett.94.086803} {\bibfield  {journal}
  {\bibinfo  {journal} {Phys. Rev. Lett.}\ }\textbf {\bibinfo {volume} {94}},\
  \bibinfo {pages} {086803} (\bibinfo {year} {2005})}\BibitemShut {NoStop}%
\bibitem [{\citenamefont {Hafezi}\ \emph {et~al.}(2007)\citenamefont {Hafezi},
  \citenamefont {S\o{}rensen}, \citenamefont {Demler},\ and\ \citenamefont
  {Lukin}}]{Lukin2007pra}%
  \BibitemOpen
  \bibfield  {author} {\bibinfo {author} {\bibfnamefont {M.}~\bibnamefont
  {Hafezi}}, \bibinfo {author} {\bibfnamefont {A.~S.}\ \bibnamefont
  {S\o{}rensen}}, \bibinfo {author} {\bibfnamefont {E.}~\bibnamefont {Demler}},
  \ and\ \bibinfo {author} {\bibfnamefont {M.~D.}\ \bibnamefont {Lukin}},\
  }\href {\doibase 10.1103/PhysRevA.76.023613} {\bibfield  {journal} {\bibinfo
  {journal} {Phys. Rev. A}\ }\textbf {\bibinfo {volume} {76}},\ \bibinfo
  {pages} {023613} (\bibinfo {year} {2007})}\BibitemShut {NoStop}%
\end{thebibliography}

\begin{thebibliography}{4}%
\makeatletter
\providecommand \@ifxundefined [1]{%
 \@ifx{#1\undefined}
}%
\providecommand \@ifnum [1]{%
 \ifnum #1\expandafter \@firstoftwo
 \else \expandafter \@secondoftwo
 \fi
}%
\providecommand \@ifx [1]{%
 \ifx #1\expandafter \@firstoftwo
 \else \expandafter \@secondoftwo
 \fi
}%
\providecommand \natexlab [1]{#1}%
\providecommand \enquote  [1]{``#1''}%
\providecommand \bibnamefont  [1]{#1}%
\providecommand \bibfnamefont [1]{#1}%
\providecommand \citenamefont [1]{#1}%
\providecommand \href@noop [0]{\@secondoftwo}%
\providecommand \href [0]{\begingroup \@sanitize@url \@href}%
\providecommand \@href[1]{\@@startlink{#1}\@@href}%
\providecommand \@@href[1]{\endgroup#1\@@endlink}%
\providecommand \@sanitize@url [0]{\catcode `\\12\catcode `\$12\catcode
  `\&12\catcode `\#12\catcode `\^12\catcode `\_12\catcode `\%12\relax}%
\providecommand \@@startlink[1]{}%
\providecommand \@@endlink[0]{}%
\providecommand \url  [0]{\begingroup\@sanitize@url \@url }%
\providecommand \@url [1]{\endgroup\@href {#1}{\urlprefix }}%
\providecommand \urlprefix  [0]{URL }%
\providecommand \Eprint [0]{\href }%
\providecommand \doibase [0]{http://dx.doi.org/}%
\providecommand \selectlanguage [0]{\@gobble}%
\providecommand \bibinfo  [0]{\@secondoftwo}%
\providecommand \bibfield  [0]{\@secondoftwo}%
\providecommand \translation [1]{[#1]}%
\providecommand \BibitemOpen [0]{}%
\providecommand \bibitemStop [0]{}%
\providecommand \bibitemNoStop [0]{.\EOS\space}%
\providecommand \EOS [0]{\spacefactor3000\relax}%
\providecommand \BibitemShut  [1]{\csname bibitem#1\endcsname}%
\let\auto@bib@innerbib\@empty
\bibitem [{\citenamefont {Tarallo}\ \emph {et~al.}(2012)\citenamefont
  {Tarallo}, \citenamefont {Alberti}, \citenamefont {Poli}, \citenamefont
  {Chiofalo}, \citenamefont {Wang},\ and\ \citenamefont {Tino}}]{Tino2012a}%
  \BibitemOpen
  \bibfield  {author} {\bibinfo {author} {\bibfnamefont {M.~G.}\ \bibnamefont
  {Tarallo}}, \bibinfo {author} {\bibfnamefont {A.}~\bibnamefont {Alberti}},
  \bibinfo {author} {\bibfnamefont {N.}~\bibnamefont {Poli}}, \bibinfo {author}
  {\bibfnamefont {M.~L.}\ \bibnamefont {Chiofalo}}, \bibinfo {author}
  {\bibfnamefont {F.-Y.}\ \bibnamefont {Wang}}, \ and\ \bibinfo {author}
  {\bibfnamefont {G.~M.}\ \bibnamefont {Tino}},\ }\href {\doibase
  10.1103/PhysRevA.86.033615} {\bibfield  {journal} {\bibinfo  {journal} {Phys.
  Rev. A}\ }\textbf {\bibinfo {volume} {86}},\ \bibinfo {pages} {033615}
  (\bibinfo {year} {2012})}\BibitemShut {NoStop}%
\bibitem [{\citenamefont {Miyake}(2013)}]{Miyake2013a}%
  \BibitemOpen
  \bibfield  {author} {\bibinfo {author} {\bibfnamefont {H.}~\bibnamefont
  {Miyake}},\ }\emph {\bibinfo {title} {{Probing and Preparing Novel States of
  Quantum Degenerate Rubidium Atoms in Optical Lattices}}},\ \href@noop {}
  {Ph.D. thesis},\ \bibinfo  {school} {Massachusetts Institute of Technology}
  (\bibinfo {year} {2013})\BibitemShut {NoStop}%
\bibitem [{\citenamefont {Kohn}(1959)}]{Kohn}%
  \BibitemOpen
  \bibfield  {author} {\bibinfo {author} {\bibfnamefont {W.}~\bibnamefont
  {Kohn}},\ }\href {\doibase 10.1103/PhysRev.115.809} {\bibfield  {journal}
  {\bibinfo  {journal} {Phys. Rev.}\ }\textbf {\bibinfo {volume} {115}},\
  \bibinfo {pages} {809} (\bibinfo {year} {1959})}\BibitemShut {NoStop}%
\bibitem [{\citenamefont {Gl\"{u}ck}\ \emph {et~al.}(2002)\citenamefont
  {Gl\"{u}ck}, \citenamefont {Kolovsky},\ and\ \citenamefont {Korsch}}]{Gluck}%
  \BibitemOpen
  \bibfield  {author} {\bibinfo {author} {\bibfnamefont {M.}~\bibnamefont
  {Gl\"{u}ck}}, \bibinfo {author} {\bibfnamefont {A.~R.}\ \bibnamefont
  {Kolovsky}}, \ and\ \bibinfo {author} {\bibfnamefont {H.~J.}\ \bibnamefont
  {Korsch}},\ }\href {\doibase http://dx.doi.org/10.1016/S0370-1573(02)00142-4}
  {\bibfield  {journal} {\bibinfo  {journal} {Physics Reports}\ }\textbf
  {\bibinfo {volume} {366}},\ \bibinfo {pages} {103 } (\bibinfo {year}
  {2002})}\BibitemShut {NoStop}%
\end{thebibliography}
\end{document}